\documentclass[10pt]{wlscirep}
\usepackage[utf8]{inputenc}
\usepackage[T1]{fontenc}
\usepackage{amsfonts}       
\usepackage{nicefrac}       
\usepackage{microtype}      
\usepackage{longtable}
\usepackage{rotating}
\usepackage{xfrac, bigints}
\usepackage{lineno, amsmath, color}
\usepackage{caption}
\usepackage{graphicx}
\usepackage{float}
\usepackage{multirow}
\usepackage{blindtext}
\usepackage{microtype}
\usepackage{pxfonts}
\usepackage{hyperref}
\allowdisplaybreaks
\title{Advanced Modeling of Lubricated Interfaces in General Curvilinear Grids}

\author[1,*]{Suhaib Ardah}
\author[2]{Francisco J. Profito}
\author[1]{Tom Reddyhoff}
\author[1]{Daniele Dini}
\affil[1]{Department of Mechanical Engineering, 
Imperial College London, London, SW7 2AZ, UK}
\affil[2]{Department of Mechanical Engineering, Polytechnic School of the University of São Paulo, São Paulo, Brazil}

\affil[*]{Corresponding author. \textit{Email address:} \texttt{s.ardah19@imperial.ac.uk}}

\begin{abstract}
Tackling fluid-flow problems involving intricate surface geometries has been the catalyst for a plethora of numerical investigations aimed at accommodating curved complex boundaries. An example is the application of body-fitted curvilinear coordinate transformation, where the one-to-one correspondence of grid points from the physical to the computational domain is achieved. In lubricated interfaces, such conversion is challenging due to the complex governing equations in the mapped-grid, the numerical instabilities exhibited by their non-linearities and the severity of operating conditions. The present contribution proposes a Reynolds-based, finite volume fluid-structure interaction (FSI) framework for solving thermal elastohydrodynamic lubrication (TEHL) problems mapped onto non-orthogonal curvilinear grids in the computational domain. We demonstrate how the strong conservation form of the pertinent governing equations can be expressed in three-dimensional curvilinear grids and discretised using finite volume method to ensure fluid-flow conservation and enforce mass-conserving cavitation conditions. Numerical and experimental benchmarks showcase the robustness and versatility of the proposed framework to simulate a diverse range of lubrication problems, hence achieving a predictive computational tool that would enable a shift towards tribology-aware design. 
\end{abstract}

\begin{document}

\keywords{Thermal elastohydrodynamic lubrication, Body-fitted curvilinear coordinates, Non-orthogonal grids, Finite volume method}

\maketitle
%
%
\thispagestyle{empty}


\section{Introduction}
The utilisation of structured rectangular grids for numerically solving Reynolds-type equations that govern the fluid flow behaviour in lubricated contacts has been used extensively due to its straightforward discretisation procedure, flexibility in imposing the boundary conditions, and lower computational effort to solve the discrete system of equations. Nevertheless, a poorly constructed grid may yield discretisation errors, particularly in problems with complex geometries, and can therefore influence the accuracy of the numerical solution. Moreover, mathematical models described in Cartesian coordinate systems may suffer from weaknesses when modelling flow at solid boundaries that are either curved or not oriented along the coordinate axes \cite{Blazek2015, Versteeg2007Feb}. Although the Reynolds model is considered to be an appropriate approximation of the full Navier-Stokes equations for solving film lubrication problems, yet the normalisation technique which traditional Reynolds-based approaches adopt aimed at imposing the hydrodynamic solution onto orthogonal rectangular grids can often result in inaccurate solutions due to the negligence of any curved features within the lubricated contact. This is more profound near the inlet and outlet regions of the contact, which are usually characterised by complex curvatures, as reported in \cite{Lee2018Sep}, where it was shown that CFD and Reynolds models yielded slightly different thermal solutions at the contact boundaries. To overcome these limitations and therefore improve the performance of fluid-structure interaction (FSI) solvers, commercial codes such as those based on the Computational Fluid Dynamics (CFD) approach have been extensively used, owing to their capabilities to capture complex geometry details in the solution of the fluid flow problem. Another strategy to deal with complicated geometries is mapping the physical flow domain onto a computational structured domain using specific coordinate transformations or body-fitted grids built via Computer-Aided Design (CAD) models \cite{Versteeg2007Feb}.

The application and development of coordinate transformation methods, particularly the technique of body-fitted curvilinear coordinate systems with coordinate lines coincident with the boundaries of arbitrarily shaped bodies, gained popularity in solving CFD problems because of their versatility in simulating flow around arbitrary physical geometries. Such coordinate transformation methods enable irregular physical domains to be mapped onto rectangular computational domains where the governing equations can be solved more efficiently, the discretisation errors minimised, and the boundary conditions imposed more accurately. Ogawa and Ishiguro \cite{Ogawa1987} solved the two-dimensional Navier-Stokes equations in body-fitted coordinates using the finite difference method to describe the incompressible flow of blood in the human ventricle and the dynamic stall process on an oscillating airfoil. Later, Yang et al. \cite{Yang1988} followed by Raithby et al. \cite{Raithby1986} studied heat transfer and fluid flow in complex geometries by applying non-orthogonal curvilinear coordinates. More recently, Cannata et al. \cite{Cannata2019} simulated wave propagation using an integral formulation of the contravariant Navier-Stokes equations in an attempt to reproduce the complex geometries found in coastal regions. Jarrah and Rizwan-Uddin \cite{JARRAH2022122559} extended the applicability of the nodal integration method scheme by deriving it in two-dimensional curvilinear coordinates to solve the convection-diffusion equation discretised by quadrilateral elements.

The implementation of the aforementioned coordinate transformation methods requires the governing equations to be written in their corresponding contravariant components form  \cite{Kajishima, Yang1994, Vinokur1974Feb, Liseikin2017Jun, Farrashkhalvat2003Mar}. However, according to an extensive review by Thompson \cite{Thompson1983} summarising the applications of grid generation techniques in fluid dynamics problems, the governing equations in the transformed domain can present additional source terms due to the differentiation of metric coefficients remaining in the gradient terms \cite{Thompson1983}. Nevertheless, such additional terms can be handled using conservative discretisation methods, such as the finite volume method (FVM), commonly used for solving fluid dynamics problems due to their inherent conservative characteristic and ability to ensure local and global flow conservation in the discrete formulation. In this case, the transformed equations written in contravariant components should be further expressed in their conversation form to properly apply the FVM. This is particularly important when the FVM is used to discretise the equations as such problem is non encountered if other discretisation methods (\emph{e.g.} finite differences and finite elements) are used, as such techniques do not require the strong conservation form of the equations, which is inherently their limitation.  Readers may refer to \cite{Versteeg2007Feb, Farrashkhalvat2003Mar, Thompson1983} for a detailed description of coordinate transformation techniques and their advantages in solving engineering applications.

To the authors’ knowledge, the derivation of the governing equations and solution of the fully coupled thermo-elastohydrodynamic lubrication problem using coordinate transformation methods has rarely been exploited. Han and Paranjpe \cite{Han1990} proposed a finite volume model to study the thermohydrodynamic performance of journal bearings using an energy equation formulated in a body-fitted coordinate system, yet unrealistic thermal boundary conditions were adopted for the bushing-oil interface, with maximum generated pressures not exceeding 100 MPa due to the conformality of the geometry. Moreover, Roberts et al. \cite{Roberts2013} implemented a finite-element-based approach for modelling multiphase lubrication problems using curvilinear shell elements; however, their analysis neglected any thermal and cavitation-related effects that may incur changes to the behaviour of the confined fluid.

The present contribution provides a new integrated formulation to solve point contact TEHL problems on a body-fitted normalised curvilinear grid. The transformation of the governing equations from physical Cartesian coordinates to non-orthogonal normalised curvilinear coordinates is achieved systematically using the Jacobian and the metric tensor of the coordinate transformation in order to convert the derivatives into the newly defined normalised curvilinear system in the computational domain. Moreover, the contravariant velocity components are derived to avoid potential numerical errors associated with the alignment of the surfaces with staggered velocity components, as suggested by \cite{Moukalled}, and to obtain the conservation forms of the transformed governing equations for the appropriated application of the finite volume method. The coupled governing equations, including the solid deformation equation, the $(p-\theta)$ generalised Reynolds equation and the fluid energy equation, are solved iteratively within an efficient and fully conservative finite volume based fluid-structure interaction (FSI) framework. The conservative aspect of the proposed framework is advantageous because it allows the more straightforward incorporation of the mass-conserving $(p-\theta)$ cavitation model in the discretisation process. Numerical benchmarks and experimental thermal data measured using infrared spectroscopy are used to validate the developed framework. Results show that the developed framework successfully predicts lubrication performance in terms of contact temperature, generated fluid pressures, traction coefficients and film thickness profiles. Although the present work focuses on counterformal contacts, the proposed method is applicable and can be easily adapted to solve any lubricated contact problem in the presence of complex geometries.

The blueprint of the current study is as follows. The governing equations of the fluid flow problem, including the $(p-\theta)$ generalised Reynolds equation, the fluid energy equation, and the fluid velocity equations, are derived in Section \ref{sec:Formulations} in the Cartesian and normalised curvilinear coordinates in their respective strong conservation forms. Section \ref{sec:Numerics_FVM} illustrates how the transformed differential equations are solved using the finite volume method, while Section \ref{sec:Framework} discusses the architecture of the FSI framework employed to solve the derived equations numerically. The validity of the developed framework is assessed in Section \ref{sec:Results} against numerical and experimental results before outlining in Section \ref{sec:Conclusion} the main findings and potential applications of the present contribution.

\section{Mathematical Formulations}
\label{sec:Formulations}
This section presents a unified derivation of the TEHL governing equations in a normalised curvilinear coordinate system that allows the solution of the full TEHL problem in a single structured computational domain while preserving the conservative nature of the equations. A concise overview of the transformation relationships between Cartesian and general curvilinear coordinates is shown in Appendix \ref{subsection:Coordinate_Transformation}. Furthermore, the general strong conservation form of the transport equation for an arbitrary scalar physical quantity written in these coordinate systems is summarised in Appendix \ref{subsection:General_Strong_Conservation_Equation}.

\begin{figure}[H]
\centering
   \includegraphics[width=0.7\linewidth]{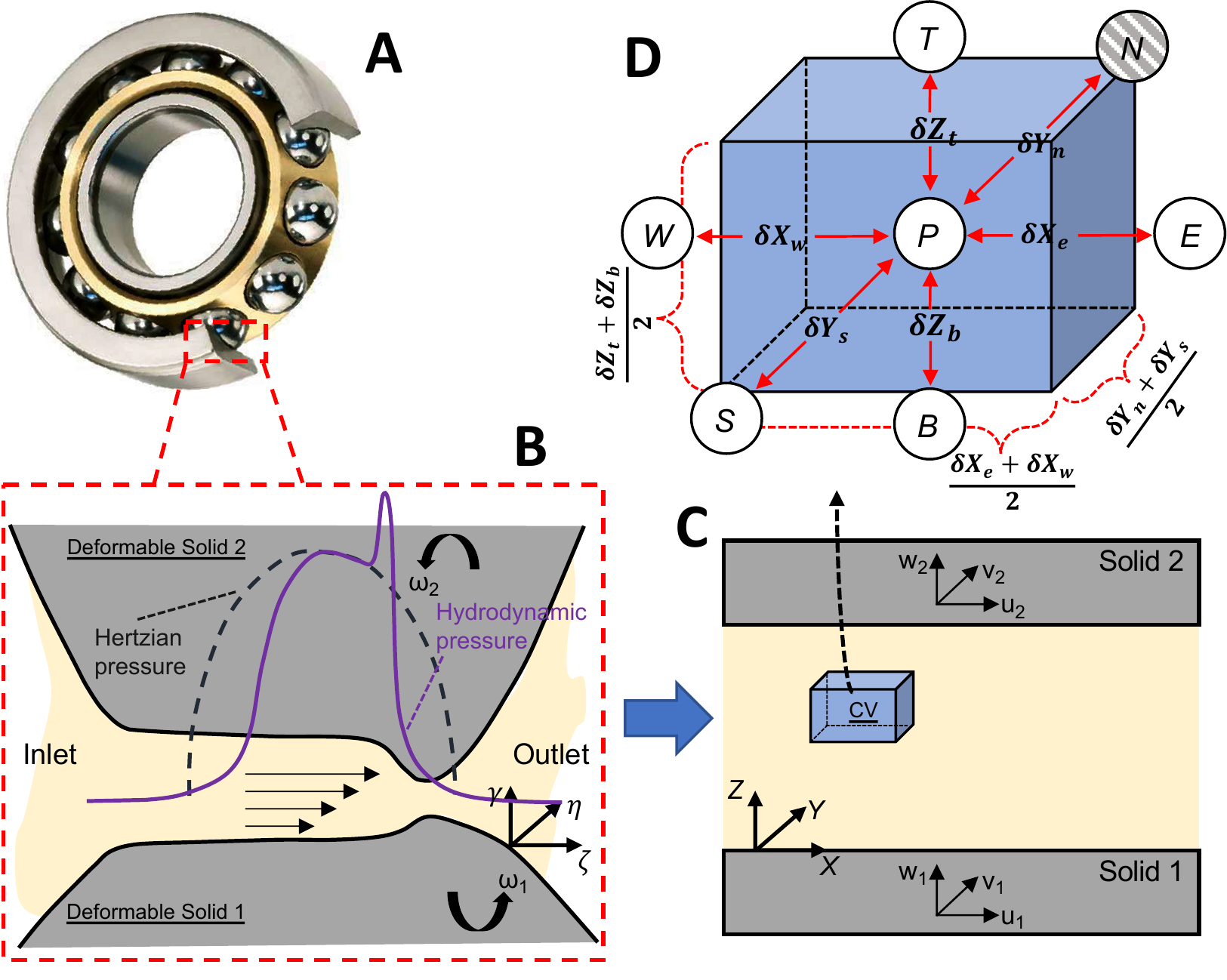}
   \caption{Characteristics of the geometry considered by the proposed framework. ($\textbf{A}$) Depicts a deep groove ball bearing, where a point contact is formed between the ball and the race, while ($\textbf{B}$) provides a magnified schematic of the contact illustrating the confined lubricant flowing in between the deformed surfaces and projected onto a curvilinear grid system ($\zeta$, $\eta$, $\gamma$). ($\textbf{C}$) Presents the transformed normalised computational domain (\emph{X}, \emph{Y}, \emph{Z}) representing the lubricated contact, and ($\textbf{D}$) elucidates the properties of the control volume (CV) with respect to the central node of the CV "P", the surrounding nodal positions and the CV faces bounding "P".}
   \label{fig:Figure_1}
\end{figure}

The proposed TEHL framework is developed based on the geometric configuration illustrated in Fig. \ref{fig:Figure_1}$\textbf{A}$, which is typical of a circular elastohydrodynamic (EHD) contact. The framework implicitly considers the dependency of the lubricant rheological properties (\emph{e.g.} dynamic viscosity, density, thermal conductivity and specific heat capacity) on temperature, pressure and shear-rate effects. However, and for the sake of brevity, the rheological models adopted to simulate the isothermal and thermal cases discussed in this study will only be reported where relevant. For the derivation of the TEHL governing equations in normalised curvilinear coordinates, the following coordinate transformation from the physical Cartesian space $x^{i} \left( x,y,z \right)$ to the mapped normalised computational space $X^{i} \left( X,Y,Z \right)$ was adopted,
\begin{equation}
  \begin{alignedat}{3}
     x & = \left( L_{x} \right) X, 
     \quad \quad \quad
     && 0 \leq x \leq L_{x} \quad && \text{and} \quad 0 \leq X \leq 1,
     \\
     y & = \left( L_{y} \right) Y = \left( \frac{L_{x}}{r_{xy}} \right) Y, 
     \quad \quad \quad
     && 0 \leq y \leq L_{y} \quad && \text{and} \quad 0 \leq Y \leq 1,
     \\
     z & = hZ + z_{1} = \left( \varepsilon L_{x} \right) \left( HZ + Z_{1} \right),
     \quad \quad \quad
     && z_{1} \leq z \leq \left( h + z_{1} \right) \quad && \text{and} \quad 0 \leq Z \leq 1,
  \label{eq:22}
  \end{alignedat}
\end{equation}
where $L_{x}$ and $L_{y}$ are the characteristic lengths of the lubricated interface in the $x$- and $y$-directions, respectively, $h(x,y)$ is the lubricant film thickness, and $z_1$ is the height of the lower surface ($z_1 = 0$ is assumed in the current study). Readers may refer to Appendix \ref{sec:Appendix_D} for a summary of the parameters and variables employed throughout this work.

\subsection{Generalised Reynolds Equation with Elrod-Adams $(p-\theta)$ Cavitation Model}
\label{subsection:GRE}
The generalised Reynolds equation (GRE) is coined as being a revamped form of the conventional Reynolds equation for considering the variation of the lubricant properties across the film thickness \cite{Dowson1962Mar}. Due to the impact of fluid-film rupture on the lubricant flow within the contact, the cavitation phenomenon is accounted for in this work by considering the Elrod–Adams $(p-\theta)$ cavitation model \cite{Elrod1975,Elrod1981} which automatically enforces the complementary Jakobsson, Floberg and Olsson (JFO) boundary conditions for cavitation \cite{Jakobsson1965, Floberg1965, Floberg1973tensile}. This entails modifying the GRE by introducing a flow parameter $\theta$ to account for the biphasic mixture within the cavitation zones \cite{Ausas2009Jul, Miraskari2017May}. Accordingly, the $(p-\theta)$ generalised Reynolds equation can be written in the strong conservation vector form in Cartesian coordinates as
\begin{align} 
    \underbrace{\boldsymbol{\nabla} \cdot \left( {\varepsilon} \boldsymbol{\nabla} p \right)}_{\text Poiseuille \; Term} = \underbrace{\boldsymbol{\nabla} \cdot \left[ \theta \left( {\rho^{*}_{e}} \boldsymbol{u_{m}} + {\rho^{*}_{1}} \boldsymbol{u_{1}} \right) \right]}_{\text Couette \; Term} \; + \underbrace{\frac{\partial \left( \theta \rho_{e} \right)}{\partial t}}_{\text Transient \; Term},
\label{eq:GRE_1}
\end{align}
where the coefficients which incorporate the variations of the lubricant properties across the film thickness are defined as
\begin{align*}
     \varepsilon  = \frac{\eta_e}{\eta^{'}_{e}} \rho^{'} - 
     \rho^{''},
     \quad \quad 
     \rho^{*}_{e} = 2 \eta_{e} \rho^{'},
     \quad \quad 
     \rho^{*}_{1} = \rho_{e} - \rho^{*}_{e},
     \quad \quad 
     \rho_{e}     = \bigintssss_{z_1}^{z_2} \rho \; dz.
\end{align*}
In the above equations, $ \boldsymbol{\nabla}(\cdot) = \left[ \dfrac{\partial }{\partial x}, \dfrac{\partial }{\partial y} \right] ^T $ is the gradient operator in Cartesian coordinates, while $ \boldsymbol{u_{m}} = \left[ u_m, v_m \right] ^T $ and $ \boldsymbol{u_{1}} = \left[ u_1, v_1 \right] ^T $ are the entrainment and lower surface velocity vectors in the $x$- and $y$-directions, respectively. The integral terms $\eta_{e}$, $\eta^{'}_{e}$, $\rho^{'}$, $\rho^{''}$ and $\rho^{*}_{e}$ are defined in Appendix \ref{sec:Appendix_A}.

The JFO boundary conditions for cavitation are preferable to other cavitation-related formulations as they impose mass conservation throughout the lubricated domain, including at the ruptured and reformation boundaries. According to this formulation, the magnitude of the fractional film content ($\theta$) is dictated by the following complementarity condition \cite{Profito2015Oct}
\begin{align} \label{eq:p-theta}
    (p - p_{cav})(1 - {\theta}) = 0 \quad {\rightarrow} \quad \left\{
    \begin{array}{lr}
     p > p_{cav} \quad {\rightarrow} \quad {\theta}=1 & \text{pressured  zones,}\\
     p = p_{cav} \quad {\rightarrow} \quad 0\;{\leq}\;{\theta}<1 &    \text{cavitated zones.}
    \end{array}\right. 
\end{align}

Considering the coordinate transformation of Eq. \ref{eq:22} in combination with the transformation relationship for spatial derivatives of Eq. \ref{eq:CoordTrans:SpatialDerivative}, along with the normalisation parameters of the dependent variables summarised in Appendix \ref{sec:Appendix_D}, the dimensionless $(p-\theta)$ generalised Reynolds equation can be expressed in the strong conservation vector form in normalised coordinates as
\begin{align} 
    {\boldsymbol{\nabla_{\scriptscriptstyle X}} \cdot \bigl( \boldsymbol{\Gamma_{D}} \boldsymbol{\nabla_{\scriptscriptstyle X}} \Bar{p} \bigr)} = \boldsymbol{\nabla_{\scriptscriptstyle X}} \cdot \bigl[ \theta \bigl( \boldsymbol{\Gamma_{CM}} \boldsymbol{U_{m}} + \boldsymbol{\Gamma_{C1}} \boldsymbol{U_{1}} \bigr) \bigr] + {\frac{\partial \left( \bar{\rho}_{e} \theta H \right)}{\partial \tau}},
\label{eq:GRE_2}
\end{align}
where
\begin{align*} 
      \boldsymbol{\Gamma_D} &= \Bar{\varepsilon} 
     \begin{bmatrix}
      1 & 0\\
      0 & r^{2}_{xy}
     \end{bmatrix},
     & \boldsymbol{\Gamma_{CM}} &= \Bar{\rho^{*}_{e}}
     \begin{bmatrix}
      H & 0\\
      0 & H r_{xy}
     \end{bmatrix},
     & \boldsymbol{\Gamma_{C1}} &= \Bar{\rho^{*}_{1}}
     \begin{bmatrix}
      H & 0\\
      0 & H r_{xy}
     \end{bmatrix}.
\end{align*}
Similarly, in the above equations, $ \boldsymbol{\nabla_{\scriptscriptstyle X} }(\cdot) = \left[ \dfrac{\partial }{\partial X}, \dfrac{\partial }{\partial Y} \right] ^T $ is the gradient operator in normalised coordinates, and $ \boldsymbol{U_{m}} = \left[ U_m, V_m \right] ^T $ and $ \boldsymbol{U_{1}} = \left[ U_1, V_1 \right] ^T $ are the dimensionless entrainment and lower surface velocity vectors in the $X$- and $Y$-directions, respectively. All relevant dimensionless parameters are found in Appendix \ref{sec:Appendix_A} and Appendix \ref{sec:Appendix_D}.

\subsection{Film Thickness Equation}
\label{subsection:FilmThickness}
The geometry of the lubricant gap in point contact EHL can be expressed in Cartesian and normalised coordinates as \cite{Zhu2019Dec}
\begin{subequations}
    \begin{align}
        h(x, y) &= h_{0} + \frac{x^{2}}{2R_x} + \frac{y^{2}}{2R_y} + {\delta}{(x,y)},
        \\
        H(X, Y) &= H_{0} + {A_{\scriptscriptstyle X} X^{2} + A_{\scriptscriptstyle Y} Y^{2}} + \bar{\delta}{(X,Y)},
    \end{align}
    \label{eq:FilmThickness}
\end{subequations}
where $h_{0}$ is the rigid separation between the contacting bodies, $R_x$ and $R_y$ are the equivalent radii curvature of the contacting solids at the centre of contact in the $x$- and $y$-directions, respectively, and $\delta$ is the pressure-induced surface normal displacement, which is calculated by the following Boussinesq convolution integral for a 3-D elastic half-space
\begin{subequations}
    \begin{align}
        {\delta}(x,y) &= \left( \frac{2}{\pi E'} \right) \bigintsss \!\!\!\!\! \bigintsss_{\Omega} \frac{{p}(x',y')}{\sqrt{\left( x - x' \right)^{2} + \left( y - y' \right)^{2}}} \,dx' dy',
        \label{eq:FilmThickness_delta_Dimensional}
        \\[5pt]
        \bar{\delta}(X,Y) &= \left( \frac{2 L_x p_0}{\pi E' h_0 r_{xy}} \right) \bigintsss \!\!\!\!\! \bigintsss_{\Bar \Omega}  \frac{\Bar{p}(X',Y')}{\sqrt{\left( X - X' \right)^{2} + {r_{xy}^{-2}} \left( Y - Y' \right)^{2}}} \,dX' dY',
        \label{eq:FilmThickness_delta_Normalised}
    \end{align}
\end{subequations}
with $E' = 2 \left[ (1-{\nu_1}^2)/E_{1}) + (1-{\nu_2}^2)/E_{2} \right] ^{-1}$ being the effective elastic modulus.

\subsection{Fluid Energy Equation}
\label{subsection:Energy_Equation}
The general energy equation for fluid film lubrication with compressible viscous fluids and internal heat transfer by diffusion governed by Fourier’s law in an isotropic medium can be written in terms of temperature in the strong conservation vector form in Cartesian coordinates as \cite{Moukalled, Habchi2018Jul}
\begin{align} \label{eq:Free_Energy_Eqn}
  \begin{split}
     \frac{\partial \left( \rho c_{p} T \right)}{\partial t} 
     + \boldsymbol{\nabla} \cdot \left( \rho c_{p}  \boldsymbol {\upsilon} T \right) 
     & = 
     \boldsymbol{\nabla} \cdot \left( k \boldsymbol{\nabla} T \right) 
     +  Q_{p} + Q_{cp} + Q_{\Phi} + \dot q_{v}, 
     \\ \\
     Q_{p} = \underbrace{{{\beta} T \left( \frac{\partial p}{\partial t} + u \frac{\partial p}{\partial x} + v \frac{\partial p}{\partial y} \right)}}_{\substack{\text{Compressive} \\ \text{Heating/Cooling}}},
     \quad \quad  
     Q_{cp} & = \underbrace{{\rho} T \left( \frac{\partial c_{p}}{\partial t} + \boldsymbol{\upsilon} \cdot \boldsymbol{\nabla} c_{p} \right)}_{\substack{\text{Enthalpic} \\ \text{Heating/Cooling}}}, \quad \quad
     Q_{\Phi} = \underbrace{{{\eta} \left[ \left( \frac{\partial u}{\partial z}\right)^2 + \left( \frac{\partial v}{\partial z}\right)^2 \right] }}_{\substack{\text{Shear} \\ \text{Heating}}},
  \end{split}
\end{align}
where $\rho$, $c_{p}$, $k$, $\beta$ and $\eta$ are the lubricant density, thermal heat capacity, thermal conductivity, thermal compressibility and dynamic viscosity, respectively. Moreover, $ \boldsymbol{\upsilon} = \left[ u, v, w \right] ^T $ is the fluid velocity vector, and $\dot q_{v}$ is the rate of heat source or sink within the fluid volume due to an arbitrary source term, such as heat released as a result of asperity-asperity contact, which is neglected in the current work.

Using the coordinate transformation relationships summarised in Appendix \ref{subsection:Coordinate_Transformation} in combination with the strong conservation form of a scalar transport equation for general non-orthogonal curvilinear coordinate systems shown in Appendix \ref{subsection:General_Strong_Conservation_Equation}, the dimensionless energy equation can be written in the strong conservation vector form in normalised coordinates as
\begingroup
\allowdisplaybreaks
\begin{gather} \label{eq:EnergyEqn_vector}
     \text{Pe} \left[ \frac{\partial \left( H \Bar{\rho} \Bar{c}_{p} \Bar{T} \right)}{\partial \tau} \right] + \text{Pe} \left[ \boldsymbol{\nabla_{\scriptscriptstyle X}} \cdot \left( \Bar{\rho} \Bar{c}_{p} \boldsymbol{\Tilde{V}} \Bar{T} \right) \right]  = \boldsymbol{\nabla_{\scriptscriptstyle X}} \cdot \left( \left[ \boldsymbol{\Bar{\Tilde{\kappa}}} \right] \boldsymbol{\nabla_{\scriptscriptstyle X}} \Bar{T} \right) + \Bar{\Tilde{Q}}_{p} + \Bar{\Tilde{Q}}_{cp} + \Bar{\Tilde{Q}}_{\Phi} + \Bar{\Tilde{\dot{q}}}_{V},
     \\[1ex]
     \begin{aligned}
         \Bar{\Tilde{Q}}_{p} &= \text{Br}^{*} \Bar{\beta} \Bar{T} \left( H \frac{\partial \Bar{p}}{\partial \tau} + \Tilde{V}^{X} \frac{\partial \Bar{p}}{\partial X} + \Tilde{V}^{Y} \frac{\partial \Bar{p}}{\partial Y} \right), \nonumber  \quad \quad 
         \Bar{\Tilde{Q}}_{cp}  = \text{Pe} \Bar{\rho} \Bar{T} \left( H \frac{\partial \Bar{c}_{p}}{\partial \tau} + \boldsymbol{\Tilde{V}} \cdot \boldsymbol{\nabla_{\scriptscriptstyle X}} \Bar{c}_{p} \right), \nonumber 
         \\[1ex]
         \Bar{\Tilde{Q}}_{\Phi} &= \frac{\text{Br} \Bar{\eta}}{H^{3}} \left[ \left( \frac{\partial \Tilde{V^{X}}}{\partial Z} \right)^{2} + \left( \frac{1}{r_{xy}} \frac{\Tilde{V}^{Y}}{\partial Z} \right)^{2}\right], \quad \quad \quad \quad 
         \Bar{\Tilde{\dot{q}}}_{V}  = H \Bar{{\dot{q}}}_{V}, \nonumber
     \end{aligned}
    \\[1ex]
    \begin{aligned}
    \left[ \boldsymbol{\bar{\Tilde{\kappa}}} \right] 
    & \approx \Bar{k}
      \begin{pmatrix}
        \varepsilon^{2} H 
        & 0 
        & - \varepsilon^{2} \left( \dfrac{\partial H}{\partial X} Z \right) 
        \\
        0 
        & \left( \varepsilon r_{xy} \right)^{2} H 
        & - \left( \varepsilon r_{xy} \right)^{2} \left( \dfrac{\partial H}{\partial Y} Z \right)
        \\
        - \varepsilon^{2} \left( \dfrac{\partial H}{\partial X} Z \right)
        & - \left( \varepsilon r_{xy} \right)^{2} \left( \dfrac{\partial H}{\partial Y} Z \right)
        & \left( \dfrac{1}{H} \right) + \varepsilon^{2} \left[ \left( \dfrac{\partial H}{\partial X} \right)^{2} + \left( r_{xy} \dfrac{\partial H}{\partial Y} \right)^{2} \right] \dfrac{Z^{2}}{H}
      \end{pmatrix} \nonumber
     \\
     & \approx \Bar{k}
      \begin{pmatrix}
        \varepsilon^{2} H 
        & 0 
        & 0 
        \\
        0 
        & \left( \varepsilon r_{xy} \right)^{2} H 
        & 0
        \\
        0
        & 0
        & \left( \dfrac{1}{H} \right) + \varepsilon^{2} \left[ \left( \dfrac{\partial H}{\partial X} \right)^{2} + \left( r_{xy} \dfrac{\partial H}{\partial Y} \right)^{2} \right] \dfrac{Z^{2}}{H}
      \end{pmatrix}, \nonumber
     \end{aligned}
\end{gather}
\endgroup
where $ \boldsymbol{\Tilde{V}} = \left[ \Tilde{V}^{X}, \Tilde{V}^{Y}, \Tilde{V}^{Z} \right] ^T $ is the dimensionless fluid velocity contravariant vector in the normalised coordinate system, and $ \left[ \boldsymbol{\Bar{\Tilde{\kappa}}} \right] $ is the dimensionless diffusion matrix. All other pertinent dimensionless parameters are found in Appendix \ref{sec:Appendix_A} and Appendix \ref{sec:Appendix_D}.  According to the lubrication theory, the magnitude of the lubricant film thickness is much smaller than the other dimensions of the contact. In this case, both the fluid pressure gradient across the film thickness and the terms of the governing equations proportional to the scaling factor $\varepsilon = \left( h_{0} / L_{x,y} \right)$ are assumed negligible, i.e., $\partial p / \partial z \approx 0$  and $\boldsymbol{\mathcal{O}}(\varepsilon^{n}) << 1$, hence why the non-diagonal elements of the diffusion matrix have been omitted. However, those neglected terms can play an important role at contact boundaries with pronounced curvatures or in regions where the scale factor $\varepsilon$ is not tiny (\emph{i.e.} at the inlet and outlet regions). Moreover, the heat convection across the film thickness and the heat diffusion along the contact (\emph{i.e.}, the first two elements of the main diagonal of $ \left[ \boldsymbol{\Bar{\Tilde{\kappa}}} \right] $) are usually neglected in lubrication problems but were kept for convenience to maintain the generality of the proposed simulation framework.

\subsection{Fluid Velocity Components}
\label{subsection:Velocity}
The Cartesian velocity components $ \boldsymbol{\upsilon} = \left[ u, v, w \right] ^T $ of the lubricant flow on the contact interface obtained in the derivation of the generalised Reynolds equation from the Navier-Stokes equation can be written as
\begin{subequations}
\begin{align}
\label{eq:Cartesian_u}
     u(x,y,z) &=
     u_{1} + \frac{\partial p}{\partial x}I_{p} + u_{s}I_{s}, \\
\label{eq:Cartesian_v}
     v(x,y,z) &=
     v_{1} + \frac{\partial p}{\partial y}I_{p} + v_{s}I_{s}, \\
\label{eq:Cartesian_w}
     w(x,y,z) &=
     - \frac{1}{\rho} \frac{\partial}{\partial x} \bigintssss_{z_{1}}^{z} \rho u \; dz' - \frac{1}{\rho} \frac{\partial}{\partial y} \bigintssss_{z_{1}}^{z} \rho v \; dz'
\end{align}
\end{subequations}
where $u_{1,2}$ and $v_{1,2}$ are the surface velocities in the $x$- and $y$-directions, respectively, while $u_s$ and $v_s$ are the sliding velocities in the respective directions. The integral coefficients $I_{p}$ and $I_{s}$ utilised in the above expressions are given in Appendix \ref{sec:Appendix_A}. Moreover, the Cartesian velocity component across the film thickness direction is calculated by integrating the mass conservation equation shown in Appendix \ref{subsection:Continuity_Equation}.

Considering the coordinate transformation relationships summarised in Appendix \ref{subsection:Coordinate_Transformation} and the normalisation parameters of the dependent variables summarised in Appendix \ref{sec:Appendix_D}, the dimensionless fluid velocity contravariant components $ \boldsymbol{\Tilde{V}} = \left[ \Tilde{V}^{X}, \Tilde{V}^{Y}, \Tilde{V}^{Z} \right] ^T $ in the normalised coordinate system can be expressed as
\begin{subequations}
 \begin{align}
     \Tilde{V}^{X} \left( X, Y, Z \right) & = H  \underbrace{\left( U_{1} + \frac{\partial \Bar{p}}{\partial X} \Bar{I}_{p} + U_{s} \Bar{I}_{s} \right)}_{U} = H U, 
     \label{eq:Ux}
     \\
     \Tilde{V}^{Y} \left( X, Y, Z \right) & = r_{xy} H  \underbrace{\left( V_{1} + r_{xy} \frac{\partial \Bar{p}}{\partial Y} \Bar{I}_{p} + V_{s} \Bar{I}_{s} \right)}_{V} = r_{xy} H V,
     \label{eq:Vx}
     \\
     \Tilde{V}^{Z} \left( X, Y, Z \right) & = W - \left( \frac{\partial H}{\partial X} U + r_{xy} \frac{\partial H}{\partial Y} V \right) Z,
     \label{eq:Wx}
 \end{align}
\end{subequations}
where the dimensionless integral coefficients $\Bar{I}_{p}$ and $\Bar{I}_{s}$ are given in Appendix \ref{sec:Appendix_A}.

\section{Numerical Solution}
\label{sec:Numerics_FVM}
This section demonstrates how the $(p-\theta)$ generalised Reynolds and fluid energy equations are solved on the mapped normalised computational domain using the finite volume method (FVM) \cite{Versteeg2007Feb, Moukalled, Ferziger2020, Chung2002Feb}. The FVM approach was adopted to numerically solve the partial differential equations (PDEs) governing the lubricant flow in TEHL contacts derived in Section  \ref{sec:Formulations} as it automatically enforces the mass flow conservation \cite{Ferziger2020}, mainly when the strong conservation form of the PDEs is applied \cite{Blazek2015}, as carried out in this study. Regardless of the chosen coordinate system, providing the strong conservation form of the governing transport equations is crucial for finite volume solutions because of the Gauss-divergence theorem that easily converts the volume integral of divergence into a surface integral of fluxes. Furthermore, the FVM has the relative advantage of being mathematically straightforward and, in particular, is excellent for problems in which quantity conservation is essential. This inherent conservative nature of the FVM is paramount in predicting the cavitation regions in lubricated contacts via the application of mass-conserving cavitation models. With regards to the Elrod-Adams $(p-\theta)$ cavitation model adopted in this work, finite volume solutions are favoured since they automatically enforce the complementary JFO boundary conditions for cavitation via the upwind-based treatment of the convective terms present in the $(p-\theta)$ generalised Reynolds equation \cite{Profito2015Oct, Ferziger2020}. Moreover, applying FVM schemes on mapped-grids has proven effective in considering the contribution of the convection and diffusion cross-derivative terms resulting from non-orthogonal coordinate transformations \cite{Demirdzic1982, Peric1982, Drikakis}.

\subsection{Finite Volume Solution of the $(p-\theta)$ Generalised Reynolds Equation}
\label{subsec: FVM_GRE}
The finite volume discretisation of the dimensionless steady-state $(p-\theta)$ generalised Reynolds equation begins by integrating Eq. \ref{eq:GRE_2} over the control volume (CV) illustrated in Fig. \ref{fig:Figure_1}D followed by applying the Gauss-divergence theorem to convert the volume integrals of divergence into surface integrals of fluxes. Accordingly:
\begin{subequations}  
 \begin{align}
      \oiiint_\mathcal{V} {\boldsymbol{\nabla_{\scriptscriptstyle X}} \cdot \left( \boldsymbol{\Gamma_{D}} \boldsymbol{\nabla_{\scriptscriptstyle X}} \Bar{p} \right)} \,d\mathcal{V} 
      &= 
      \oiiint_\mathcal{V} \boldsymbol{\nabla_{\scriptscriptstyle X}} \cdot \bigl[ \theta \bigl( \boldsymbol{\Gamma_{CM}} \boldsymbol{U_{m}} + \boldsymbol{\Gamma_{C1}} \boldsymbol{U_{1}} \bigr) \bigr]  \,d\mathcal{V},
      \\
     \oiint_{\partial \mathcal{S}} \left( \boldsymbol{\Gamma_{D}} \boldsymbol{\nabla_{\scriptscriptstyle X}} \Bar{p} \right) \cdot \vec{\boldsymbol{n}} \,d\mathcal{S} 
     &= 
     \oiint_{\partial \mathcal{S}} \bigl[ \theta \bigl( \boldsymbol{\Gamma_{CM}} \boldsymbol{U_{m}} + \boldsymbol{\Gamma_{C1}} \boldsymbol{U_{1}} \bigr) \bigr] \cdot \vec{\boldsymbol{n}} \,d\mathcal{S},
     \label{eq:AA}
 \end{align}
\end{subequations}
where $\mathcal{V}$ is the CV volume of unit height (since the dimensionless GRE is two-dimensional defined in the $XY$-plane) enclosing node $P$ as shown in Fig. \ref{fig:Figure_1}D, and $\vec{\boldsymbol{n}}$ is the unit vector normal to the CV faces $\mathcal{S}$. The integrals over the CV faces can be approximated using the midpoint rule such that the diffusive and convective fluxes are evaluated at the centroid of each face \cite{Moukalled, Profito2015Oct, Ferziger2020}. This entails expressing the surface integrals as a summation of the diffusive and convective fluxes travelling through the CV faces, knowing that the surface unit vectors on opposite sides of the CV have opposite signs. Therefore, the discrete form of Eq. \ref{eq:AA} formulated in terms of the central CV node ($P$) and the surrounding CV faces ($f$) becomes
\begin{align}  
      \sum_{f \sim nb(P)} \left( \boldsymbol{\Gamma_{D}} \boldsymbol{\nabla_{\scriptscriptstyle X}} \Bar{p}  \right)_{f} \cdot \boldsymbol{\mathcal{S}}_{f}
      =
      \sum_{f \sim nb(P)} \bigl[ \theta \bigl( \boldsymbol{\Gamma_{CM}} \boldsymbol{U_{m}} + \boldsymbol{\Gamma_{C1}} \boldsymbol{U_{1}} \bigr) \bigr]_{f} \cdot \boldsymbol{\mathcal{S}}_{f},
\end{align}
where $\boldsymbol{\mathcal{S}}_{f} = ( {S_f}\vec{\boldsymbol{n}}_f )$ represents the surface vectors normal to the CV faces of unit heights, which include the orientation of the fluxes and the face lengths/widths they pass through. The diffusive and convective fluxes at the CV faces were approximated using the central (CDS) and upwind (UWD) differencing schemes, respectively, based on the quantities of the neighbouring CV centroids. 

The semi-system approach proposed initially by \cite{Ai1993} is utilised in the present work to improve the stability of the iterative numerical solution of the EHL problem. This is carried out by expressing the film thickness in the Couette term (R.H.S) of the GRE at a nodal position ($i,j$) as a function of the neighbouring nodal pressures using the influence coefficients $ \left( D^{i,j}_{k,l} \right)$ \cite{Wang2020Aug, Hough1929Jan} given by the discretisation of the Boussinesq integral and assuming a piecewise zero-order polynomial pressure distribution, as shown in Eq. \ref{eq:CC}. Such an approach ensures the diagonal dominance of the coefficient matrix in the solution of the GRE, ergo avoiding an ill-conditioned matrix which stabilises the iterative solution of the linear system of equations, particularly in thin film problems or under severe loading conditions where fluid viscosity could rise to several orders of magnitude.
\begin{align}
     H_{i, j} 
     = H_{0} + \frac{X_{i}^{2} + Y_{j}^{2}}{2R'} + \biggl[ D^{i,j}_{k-1,l} \left( \Bar{p}_{k-1,l} \right) + D^{i,j}_{k,l} \left( \Bar{p}_{k,l} \right) + D^{i,j}_{k+1,l} \left( \Bar{p}_{k+1,l} \right) \biggr].
     \label{eq:CC}
\end{align}

Therefore, the linear system of equations derived from the finite volume discretisation of the dimensionless $(p-\theta)$ generalised Reynolds equation for semi-system EHL solution can be written as
\begin{align} 
     \boldsymbol{\alpha} \left( \Bar{p}_{W} \right) + \boldsymbol{\beta} \left( \Bar{p}_{P} \right) + \boldsymbol{\gamma} \left( \Bar{p}_{E} \right) + \boldsymbol{\zeta} \left( \Bar{p}_{N} \right) + \boldsymbol{\lambda} \left( \Bar{p}_{S} \right) = \boldsymbol{\varphi},
     \label{eq:Discrete_GRE}
\end{align}
where $\boldsymbol{\alpha}$, $\boldsymbol{\beta}$, $\boldsymbol{\gamma}$, $\boldsymbol{\zeta}$, $\boldsymbol{\lambda}$ and $\boldsymbol{\varphi}$ are defined in Appendix \ref{sec:Appendix_B}.

\subsection{Finite Volume Solution of the Fluid Energy Equation}
The finite volume discretisation of the dimensionless steady-state energy equation for fluid film lubrication follows the same steps adopted for the GRE, except that for the fluid energy equation, the discretisation is conducted in a three-dimensional grid. Hence, by integrating Eq. \ref{eq:EnergyEqn_vector} over the control volume (CV) illustrated in Fig. \ref{fig:Figure_1}D followed by applying the Gauss-divergence theorem one has
\begin{subequations}  \label{eq:Discrete_EnergyEqn}
 \begin{align}
      \text{Pe} \oiiint_\mathcal{V} \left[ \boldsymbol{\nabla_{\scriptscriptstyle X}} \cdot \left( \Bar{\rho} \Bar{c}_{p} \boldsymbol{\Tilde{V}} \Bar{T} \right) \right] \,d\mathcal{V} &= \oiiint_\mathcal{V} \boldsymbol{\nabla_{\scriptscriptstyle X}} \cdot \left( \left[ \boldsymbol{\Bar{\Tilde{\kappa}}} \right]^{T} \boldsymbol{\nabla_{\scriptscriptstyle X}} \Bar{T} \right) \,d\mathcal{V} + \oiiint_\mathcal{V} \Bar{\Tilde{Q}}_{T} \,d\mathcal{V},
      \\
      \text{Pe} \oiint_{\partial \mathcal{S}} \left( \Bar{\rho} \Bar{c}_{p} \boldsymbol{\Tilde{V}} \Bar{T} \right) \cdot \vec{\boldsymbol{n}} \,d\mathcal{S} &= \oiint_{\partial \mathcal{S}} \left( \left[ \boldsymbol{\Bar{\Tilde{\kappa}}} \right]^{T} \boldsymbol{\nabla_{\scriptscriptstyle X}} \Bar{T} \right) \cdot \vec{\boldsymbol{n}} \,d\mathcal{S} + \oiiint_{\mathcal{V}} \Bar{\Tilde{Q}}_{T} \,d\mathcal{V}.
      \label{eq:FVM_SurfaceArea}
 \end{align}
\end{subequations}

In addition to the midpoint rule for approximating the flux integrals at the centroid of each CV face, the source term can be approximated as the product between its magnitude at the CV centre and the CV volume. Hence \cite{Moukalled}:
\begin{align}  
      \text{Pe} \sum_{f \sim nb(P)} \left( \Bar{\rho} \Bar{c}_{p} \boldsymbol{\Tilde{V}} \Bar{T} \right)_{f} \cdot \boldsymbol{\mathcal{S}}_{f}
      =
      \sum_{f \sim nb(P)} \left( \left[ \boldsymbol{\Bar{\Tilde{\kappa}}} \right]^{T} \boldsymbol{\nabla} \Bar{T}  \right)_{f} \cdot \boldsymbol{\mathcal{S}}_{f} + \left( \Bar{\Tilde{Q}}_{T} \right)_{P} \cdot \left( \Delta \mathcal{V} \right)_{P},
\end{align}
where $\Delta \mathcal{V}$ is the CV volume whereas $\boldsymbol{\mathcal{S}}_{f}$ is the surface vectors normal to the CV faces that include the orientation of the fluxes and the face areas they pass through. The convective and diffusive fluxes were also approximated using the upwind and central differencing schemes, respectively. Therefore, the system of algebraic equations derived from the finite volume discretisation of the dimensionless fluid energy equation can be written as
\begin{align}  
    \boldsymbol{\alpha} \left( \Bar{T}_{W} \right) + \boldsymbol{\beta} \left( \Bar{T}_{P} \right) + \boldsymbol{\gamma} \left( \Bar{T}_{E} \right) + \boldsymbol{\zeta} \left( \Bar{T}_{N} \right) + \boldsymbol{\lambda} \left( \Bar{T}_{S} \right) + \boldsymbol{\phi} \left( \Bar{T}_{T} \right) + \boldsymbol{\psi} \left( \Bar{T}_{B} \right) = \boldsymbol{\varphi},
    \label{eq:Discrete_Energy}
\end{align}
where the above coefficients are defined in Appendix \ref{sec:Appendix_C}.

\subsection{Fluid-Structure Interaction Framework}
\label{sec:Framework}
The FSI framework developed by the authors in \cite{Ardah2023Jan} for simulating two-dimensional TEHL problems has been scaled up in this work to accommodate three-dimensional TEHL solutions following the proposed mathematical and numerical formulations described in the previous sections. The framework begins by initialising the EHL solver to calculate the lubricant pressure and film thickness while neglecting temperature changes. Afterwards, the results of the EHL solver are fed into the TEHL solver to update the temperature distributions in the lubricant film and solids. The overall convergence is obtained when relative pressure and thermal solutions errors are minimised below a certain convergence criterion. Although a brief description of the EHL and TEHL solvers will be provided here, readers may refer to Section 4.1 from \cite{Ardah2023Jan} for a detailed description of the simulation framework.

The EHL solver calculates the lubricant pressure and film thickness through an iterative process. Using an iterative Gauss-Seidel line relaxation method, the lubricant pressure $\left( \Bar p \right)$ and film fraction $\left( \theta \right)$ are computed by solving the dimensionless $(p-\theta)$ generalised Reynolds equation (Eq. \ref{eq:Discrete_GRE}). Concurrently, lubricant properties, such as viscosity $\left( \Bar \eta \right)$ and density $\left( \Bar \rho \right)$, are updated due to pressure and shear-rate effects using pre-defined rheological models. At the same time, the pressure-induced surface deformation $\left( \bar{\delta} \right)$ and its influence on film thickness $\left( H \right)$ are evaluated using Eq. \ref{eq:FilmThickness}, where the discrete convolution and fast Fourier transform (DC-FFT) algorithm \cite{Wang2020Aug, Liu2000Aug} are adopted to solve the Boussinesq integral efficiently. The EHL iterative process terminates when the relative error between the computed pressures $\left( \Bar{p} \right)$ at successive iterations is minimised, and the equilibrium of forces over the contact domain ${\Omega}$, given in the dimensionless form in Eq. \ref{eq:EquilibriumForces} for point contact problems, is achieved.
\begin{align}  \label{eq:EquilibriumForces}
    \iint_{\Omega} \Bar{p}(X, Y) \,dX \,dY = \frac{2 \pi}{3}.
\end{align}

Subsequently, the converged EHL results initialise the thermal loop, simultaneously calculating the lubricant $\left( \Bar T \right)$ and solids temperatures $\left( \Bar{T}_{s{1,2}} \right)$ iteratively and updating the fluid rheological behaviour. The dimensionless fluid energy equation (Eq. \ref{eq:Discrete_Energy}) is solved using a Gauss-Seidel line relaxation, whereas the solids temperatures are calculated similarly but neglecting any source term. The coupling between the thermal-fluid and thermal-solid solvers is established by imposing heat flux continuity at the fluid-solid interfaces using the conjugate heat transfer method \cite{Verstraete2016Oct}. The numerical convergence of the thermal loop is achieved when the errors of the lubricant and solid temperatures, the heat flux transmitted and the temperatures at the fluid-solid interfaces are minimised below a specified tolerance between successive iterative steps.

Once the EHL and thermal solvers converge, a final convergence check on the pressure and temperatures is performed to evaluate the global convergence of the TEHL solution. The EHL solver is then recalled if global convergence is not achieved, thus repeating the steps mentioned above until the global convergence is reached. Point Gauss-Seidel Method with Aitken Acceleration (PGMA) is integrated into the simulation framework to improve the convergence of the iterative solvers. Aimed at coupling black-box solvers in fluid-structure interaction problems \cite{Degroote2013Sep}, PGMA is based on the dynamic variation of the relaxation parameters and has proved to be more effective in simulating two-dimensional TEHL problems than Fixed Point Gauss-Seidel Method (PGMF), where the latter employs constant relaxation parameters \cite{Ardah2023Jan}. The PGMA method updates the magnitude of the relaxation factors (${\omega_{\psi}^{n}}$) for an arbitrary field variable $\psi$ using residues of the current $\left( \vec{\boldsymbol{\mathfrak{R}}}_{\psi}^{n} \right)$ and previous $\left( {\vec{\boldsymbol{\mathfrak{R}}}_{\psi}^{n-1}} \right)$ iterations as follows
\begin{subequations}
 \begin{align}
  \begin{split}
     {\omega_{\psi}^{n}} 
     &= {\omega_{\psi}^{n-1}} \left[ \frac{\left( {\vec{{\boldsymbol{\mathfrak{R}}}}_{\psi}^{n-1}} \right)^{{T}}\cdot\left( {\vec{{\boldsymbol{\mathfrak{R}}}}_{\psi}^{n-1}} - {\vec{{\boldsymbol{\mathfrak{R}}}}_{\psi}^{n}} \right)}{{\left( {\vec{{\boldsymbol{\mathfrak{R}}}}_{\psi}^{n}} - {\vec{{\boldsymbol{\mathfrak{R}}}}_{\psi}^{n-1}} \right)}^{{T}}\cdot\left( {\vec{{\boldsymbol{\mathfrak{R}}}}_{\psi}^{n}} - {\vec{{\boldsymbol{\mathfrak{R}}}}_{\psi}^{n-1}} \right)} \right],
     \\[1ex]
     {\omega_{\psi}^{n}} \; 
     &{\xleftarrow \; \text{min}\left[\text{max}\left({\omega_{\psi}^{n}}, \; {\omega_{min}} \right), \; {\omega_{max}} \right]},
  \end{split}
 \end{align}
where $\omega_{min}$ and $\omega_{max}$ are usually set as 0.001 and 1.0, respectively, and are imposed to limit the magnitude of ${\omega_{\psi}^{n}}$. The residue and the update of a field variable in a given iteration $n$ are calculated as
 \begin{align}
  \begin{split}
    {\vec{\boldsymbol{\mathfrak{R}}}^{n}_{\psi}} &= {\vec{\boldsymbol{\psi}}^{n}} - {\vec{\boldsymbol{\psi}}^{n-1}},
    \\[1ex]
    {\vec{\boldsymbol{\psi}}^{n}} &= {\vec{\boldsymbol{\psi}}^{n-1}} + \omega_{\psi}^{n} \left( {\vec{\boldsymbol{\mathfrak{R}}}^{n}_{\psi}} \right).     
  \end{split}
 \end{align}
\end{subequations}
Readers may refer to \cite{Ardah2023Jan} for a detailed description of PGMA implementation and a comprehensive analysis of the improved convergence rates that the PGMA offers compared to the PGMF partitioned technique.

\section{Results and Discussion}
\label{sec:Results}
The accuracy of the developed TEHL simulation framework is assessed using numerical results representative of thermal point-loaded contacts found in the literature. Furthermore, temperature measurements obtained using infrared spectroscopy and experimental traction data are analysed to corroborate the proposed framework's correctness further. The rheological models used to characterise the lubricant behaviour due to changes in pressure, temperature and shear-rate are specified according to each relevant study considered.

\subsection{Numerical Benchmark}
The first numerical benchmark is based on a series of investigations carried out by Kaneta et al. \cite{Kaneta2014Dec} aimed at studying the influence of the slide-roll ratio (SRR) on the lubrication performance of a $\mathrm{Si_{3}N_{4}}$-steel contact. The oil viscosity change due to pressure and temperature effects is described using the thermal Roelands model (Eq. \ref{eq:mu_Kaneta_PHT}) and the Eyring model (Eq. \ref{eq:nonNewtonian_Kaneta_PHT}) is adopted to account for the lubricant non-Newtonian behaviour, while the oil density is calculated using the Dowson and Higginson relationship (Eq. \ref{eq:rho_Kaneta_PHT}). All the parameters and operating conditions required for the TEHL simulation can be found in \cite{Kaneta2014Dec}.
\begin{subequations}
    \begin{align}
        \eta &= \eta_0 \; \exp \Biggl\{ ( \ln \eta_0 + 9.67 ) \times \bigg[ -1 + \left( 1 + 5.1 \times 10^{-9} p \right)^{\dfrac{\alpha}{5.1 \times 10^{-9} \left( \ln \eta_{0} + 9.67 \right)}} \left( \frac{T - 138}{T_0 - 138} \right)^{- \dfrac{\beta (T_{0} - 138)}{\left( \ln \eta_{0} + 9.67 \right)}} \bigg] \Biggl\},
        \label{eq:mu_Kaneta_PHT}
        \\[1ex]
        \eta^{*} &= \frac{\eta \left( \tau_{e} / \tau_{0} \right)} {\sinh{\left( \tau_{e} / \tau_{0} \right)}},
        \label{eq:nonNewtonian_Kaneta_PHT}
        \\[1ex]
        \rho &= \rho_{0} \left( 1 + \frac{0.6 \times 10^{-9} p}{1 + 1.7 \times 10^{-9} p} \right) \left[ 1 - 0.00065 \left( T - T_0 \right) \right].
        \label{eq:rho_Kaneta_PHT}
    \end{align}
\end{subequations}

\begin{figure}[H]
\centering
   \includegraphics[width=0.85\linewidth]{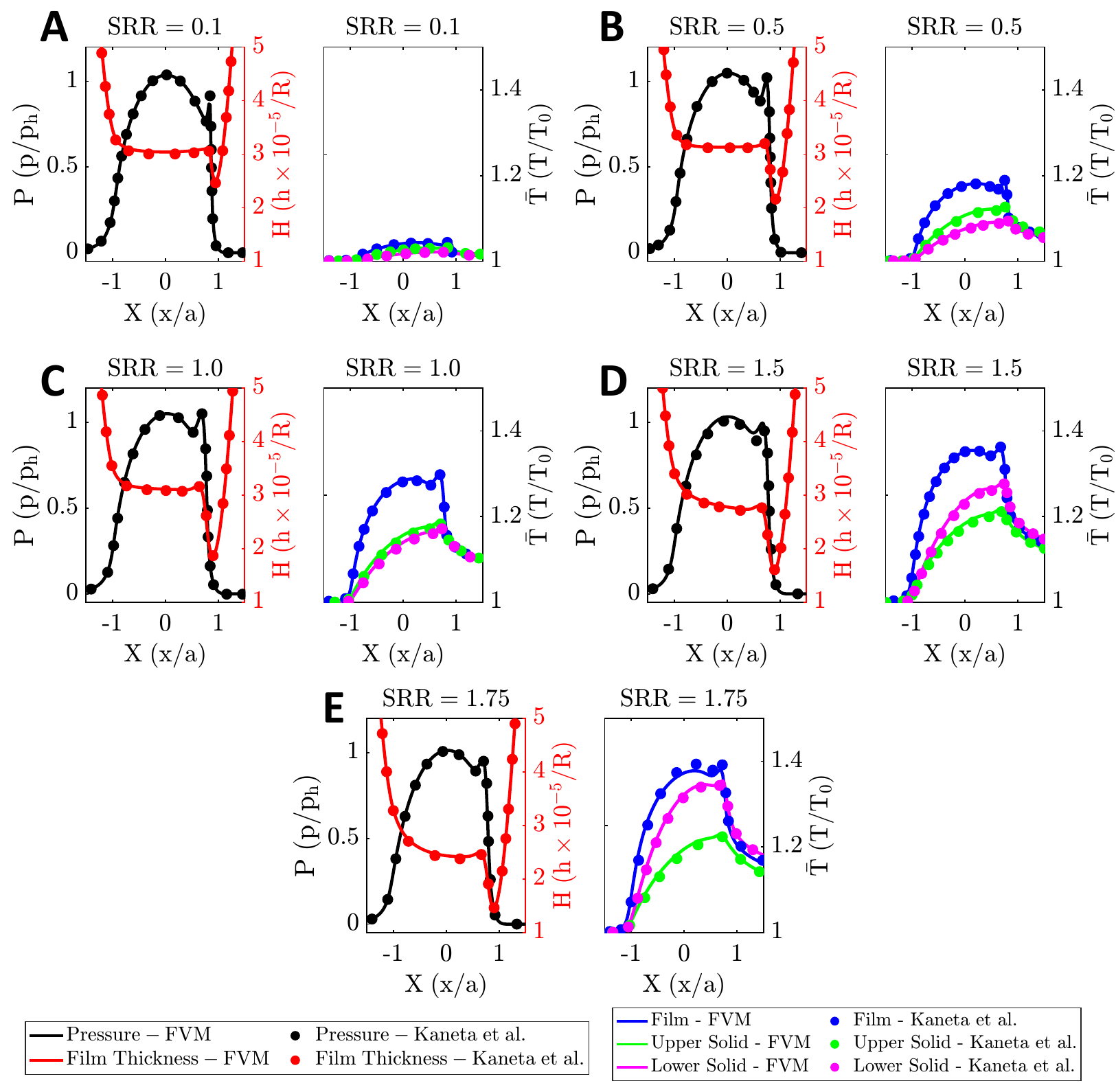}
   \caption{Comparison of the mid-plane ($Y=0$) dimensionless pressure and film thickness profiles and lubricant film, steel (lower solid) and $\mathrm{Si_{3}N_{4}}$ (upper solid) surface temperatures along the entertainment $X$-direction calculated with the proposed TEHL framework (lines) against their counterparts (dots) in Kaneta et al. \cite{Kaneta2014Dec}. The cases are representative of $p_h = 1 \; \text{GPa}$, $R^{'} = 0.02 \; \text{m}$, $T_0 = 303 \; \text{K}$, $u_e = 1.32 \; \text{m/s}$ and slide-roll-ratios of (\textbf{A}) SRR = 0.1, (\textbf{B}) SRR = 0.5, (\textbf{C}) SRR = 1.0, (\textbf{D}) SRR = 1.5 and (\textbf{E}) SRR = 1.5.}
   \label{fig:Kaneta_PHT}
\end{figure}

It can be deduced from Fig. \ref{fig:Kaneta_PHT} that a good agreement is achieved for pressure, film thickness and temperature profiles. As verified in the figure, the film temperature increases with SRR due to the higher heat generated by shear near the mid-plane of the film, thus explaining the higher lubricant temperature compared to the surface temperatures regardless of the SRR. Because of the temperature rise in the contact, fluid viscosity tends to decrease, resulting in lower film thickness for the highest SRRs. Furthermore, the difference in the surface temperatures can be explained by the different thermal conductivities of the surfaces, wherein the higher temperatures are found for the $\mathrm{Si_{3}N_{4}}$ surface ($k = 23 \; \text{W/mK}$) as it carries less heat away compared to steel ($k = 46 \; \text{W/mK}$). This is the case for low SRR, while the situation differs for higher SRRs where convection effects dominate.

The second numerical benchmark employed to validate the developed TEHL simulation framework is based on another study conducted by Kaneta et al. \cite{Kaneta2022Jun} aimed at assessing the thermal behaviour of lubricated contacts when solids of dissimilar thermal properties are considered. The point contact configurations are that of steel-SiC, steel-$\mathrm{Al_{2}O_{3}}$, and steel-$\mathrm{Si_{3}N_{4}}$ contacts where each solid pair is separated by a continuously variable transmission (CVT) fluid and loaded with a maximum Hertizan pressure $p_h = 0.75 \, \text{GPa}$. The input parameters required for the TEHL simulations, such as the thermophysical properties of the solids and lubricant, can be found in \cite{Kaneta2022Jun}. The viscosity- and density-pressure-temperature dependencies of the lubricant are described via the following Ohno's relationships \cite{Ohno2007Feb}
\begin{subequations}
    \begin{align}
        \log_{10} \eta &= 7 - \frac{11.3 \left( T - T_{VE} \right) \left( \frac{206}{T_{VE}} \right)} {35.9 + \left( T - T_{VE} \right) \left( \frac{206}{T_{VE}} \right)},
        \label{eq:mu_Kaneta_TU}
        \\[5pt]
        \ln \rho &= \ln \rho_{0} + 0.0012T \left( 1 - \frac{206}{T_{VE}} \right) + \frac{p}{12.8 \times 10^8},
        \label{eq:rho_Kaneta_TU}
    \end{align}
\end{subequations}
where $T_{VE} = 206 + 209.4 \ln \left( 1 + 1.445 \times 10^{-9} p \right)$, while the non-Newtonian lubricant behaviour is described using the Eyring shear-thinning model given in Eq. \ref{eq:nonNewtonian_Kaneta_PHT}.

\begin{figure}[H]
\centering
   \includegraphics[width=0.75\linewidth]{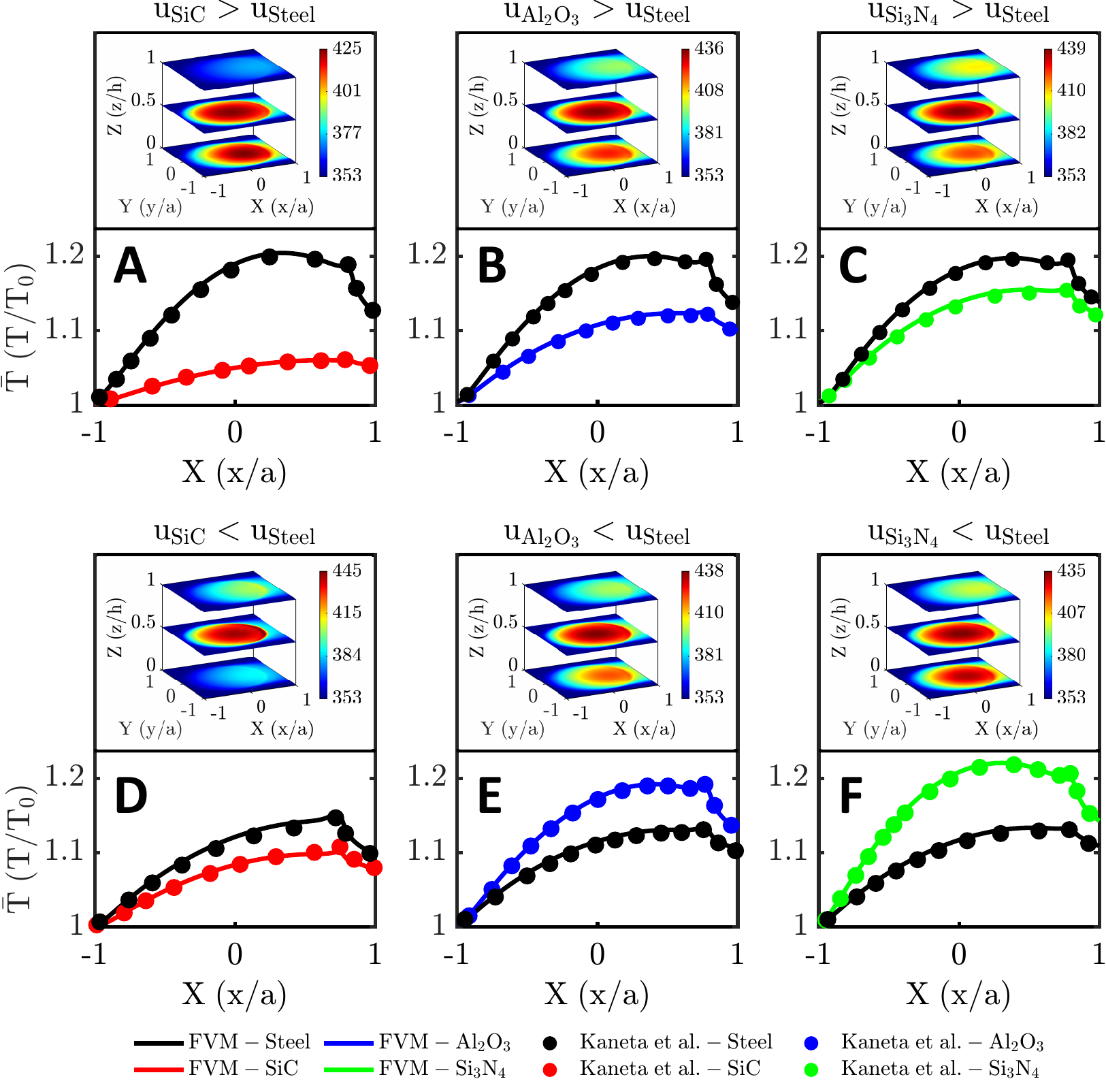}   
   \caption{Comparison of the mid-plane ($Y=0$) dimensionless surface temperature profiles along the entertainment $X$-direction calculated with the proposed TEHL framework (lines) against their counterparts (dots) in \cite{Kaneta2022Jun}. (\textbf{A}), (\textbf{B}) and (\textbf{C}) portray the surface temperatures when the SiC, $\mathrm{Al_{2}O_{3}}$ and $\mathrm{Si_{3}N_{4}}$ surfaces are travelling faster than the steel surface, respectively, while (\textbf{D}), (\textbf{E}) and (\textbf{F}) correspond to the thermal results during which the steel surface is faster than the SiC, $\mathrm{Al_{2}O_{3}}$ and $\mathrm{Si_{3}N_{4}}$ surfaces, respectively. Dimensional thermal maps (in Kelvin) pertinent to the bounding surfaces and mid-layer of the confined film are illustrated for each contact pair and speed condition. The numerical investigations were carried out for $p_h = 0.75 \, \text{GPa}$, $T_0 = 353 \, \text{K}$, SRR = 1 and entrainment speeds of $u_{e}|_{SiC-Steel} = 4.4 \; \text{m/s}$, $u_{e}|_{\mathrm{Al_{2}O_{3}}-Steel} = 4.1 \; \text{m/s}$ and $u_{e}|_{\mathrm{Si_{3}N_{4}}-Steel} = 3.8 \; \text{m/s}$.}
   \label{fig:Kaneta_TU_1} 
\end{figure}

\begin{figure}[H]
\centering
   \includegraphics[width=\linewidth]{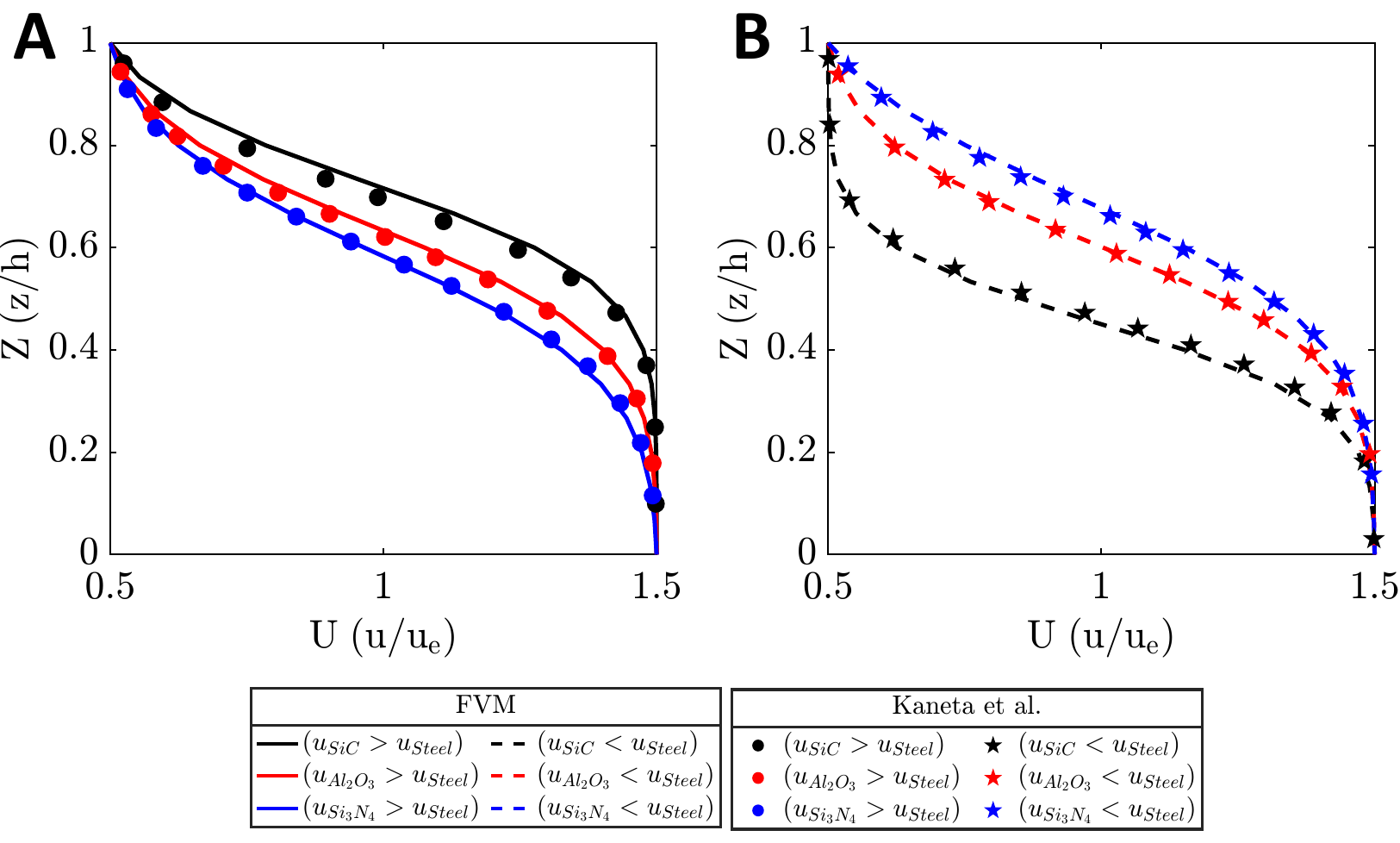}
   \caption{Dimensionless lubricant velocity profiles across the film thickness at ($X=Y=0$) calculated with the proposed TEHL framework (lines) against their counterparts (dots) in \cite{Kaneta2014Dec} for SRR = 1 and entrainment speeds of $u_{e}|_{SiC-Steel} = 4.4 \; \text{m/s}$, $u_{e}|_{\mathrm{Al_{2}O_{3}}-Steel} = 4.1 \; \text{m/s}$ and $u_{e}|_{\mathrm{Si_{3}N_{4}}-Steel} = 3.8 \; \text{m/s}$.}
   \label{fig:Kaneta_TU_2} 
\end{figure}

\begin{figure}[H]
\centering
   \includegraphics[width=\linewidth]{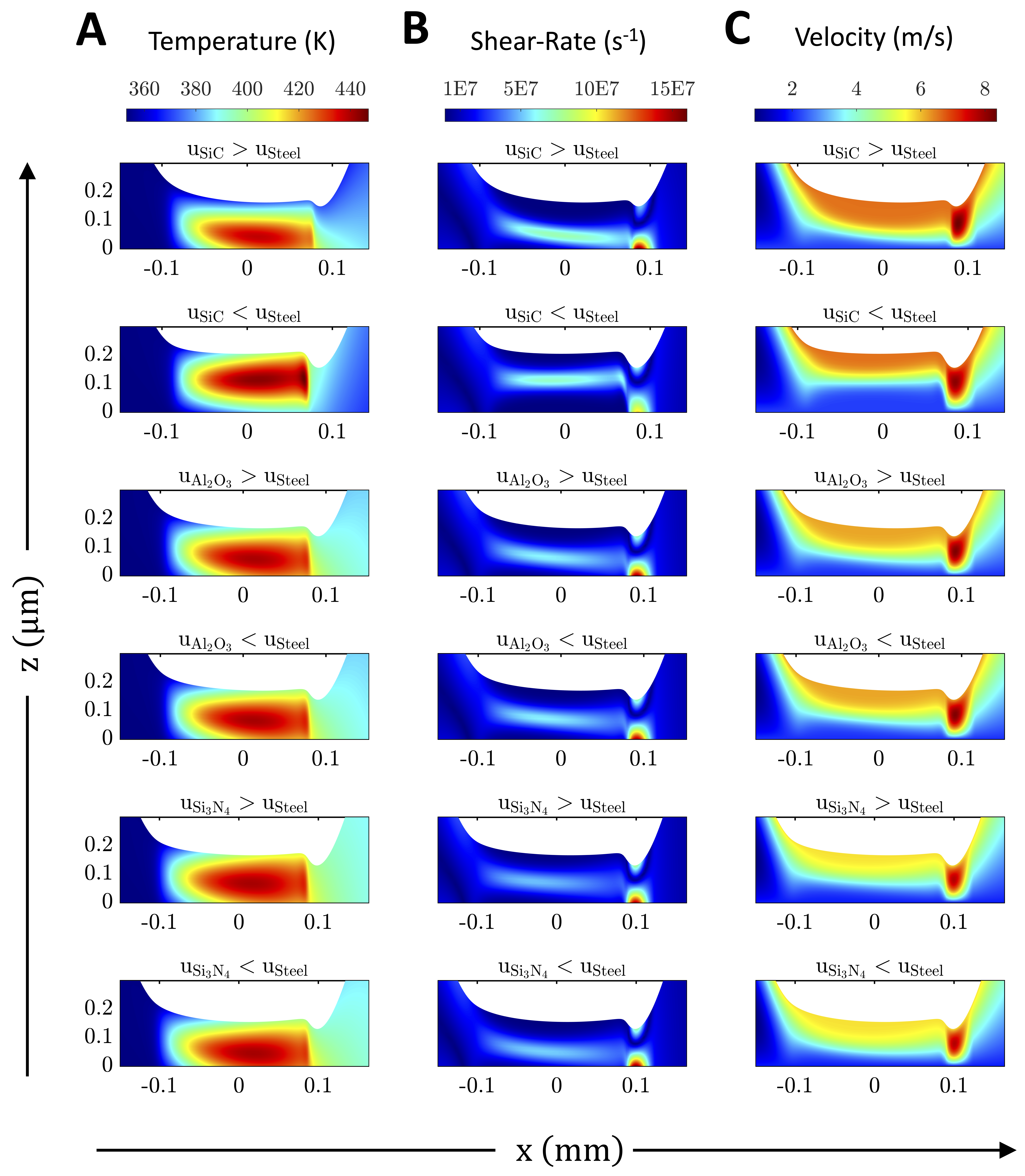}   
   \caption{Mid-plane ($Y=0$) contour maps corresponding to the operating conditions in Figs. \ref{fig:Kaneta_TU_1} and \ref{fig:Kaneta_TU_2}. \textbf{(A)} Lubricant temperature, \textbf{(B)} shear-rate and \textbf{(C)} lubricant velocity distributions.}
   \label{fig:Kaneta_TU_3} 
\end{figure}

The robustness of the developed TEHL simulation framework can be inferred from the comparisons illustrated in Fig. \ref{fig:Kaneta_TU_1}. Inevitably, the predicted temperature of the SiC surface is always higher than that of steel regardless of how fast the solids move relative to each other due to the significant difference in thermal conductivity of the two materials ($k_{SiC} = 200 \, \text{W/mK}$ while $k_{steel} = 21 \, \text{W/mK}$). On the other hand, switching the surface velocity directions has little influence on the magnitude of the maximum solid temperatures in the case of steel-$\mathrm{Al_{2}O_{3}}$ contact, where the thermal conductivity of $\mathrm{Al_{2}O_{3}}$ ($k_{\mathrm{Al_{2}O_{3}}} = 29 \, \text{W/mK}$) is similar to that of steel; however, higher temperatures are always found on the slower surface, which carries less heat away from the contact, as depicted in Figs. \ref{fig:Kaneta_TU_1}B and \ref{fig:Kaneta_TU_1}E. Although the thermal conductivity of steel and $\mathrm{Si_{3}N_{4}}$ are almost identical ($k_{\mathrm{Si_{3}N_{4}}} = 23 \, \text{W/mK}$), yet it is evident from Figs. \ref{fig:Kaneta_TU_1}C and \ref{fig:Kaneta_TU_1}F that a significant change in solid temperatures can be observed depending on which surface is travelling at a higher speed. To further corroborate the proposed TEHL framework, Fig. \ref{fig:Kaneta_TU_2} portrays how the dimensionless lubricant velocity profiles across the film thickness at ($X=Y=0$) calculated using Eqs.\ref{eq:Ux}, \ref{eq:Vx} and \ref{eq:Wx} agree very well with those evaluated in \cite{Kaneta2022Jun} under thermal conditions. Furthermore, to gain a more quantitative understanding of those results, Fig. \ref{fig:Kaneta_TU_3} illustrates the mid-plane ($Y=0$) temperature, shear-rate and lubricant velocity distributions for the different pair of materials.

\newpage
\subsection{Experimental Benchmark}
In addition to validating against numerical-based results, the use of experimental data enhances the reliability of the proposed set of governing equations and evaluates its accuracy in predicting performance of lubricated contacts. Thermal and traction results measured in an experimental study carried out by Reddyhoff et al. \cite{Lu2020Jun} are used in the current work. It is important to note that the use of temperature data for corroboration of numerical results is unconventional and that the majority of validation exercises are often limited to friction and film thickness measurements. The authors in \cite{Lu2020Jun} determined the average temperature of a ball and disk point contact using infrared spectroscopy technique where emissivity radiated from the oil, ball and disk is captured and isolated using band pass optical filters and a chromium coating applied to the disk. The obtained radiation is then converted to thermal data according to Planck's Law and several calibration tests at different temperature and film thickness \cite{Lu2020Jun}. The tests were conducted on a steel ($k_{\mathrm{Steel}} = 21 \, W/m.K$), silicon nitride ($k_{\mathrm{Si_{3}N_{4}}} = 30 \, W/m.K$) and zirconia ($k_{\mathrm{Zirconia}} = 3 \, W/m.K$) ball loaded against a sapphire disk and separated using the traction fluid Santotrac 50 and PAO 4. Operating conditions such as applied load and entrainment speeds were varied during the tests \cite{Lu2020Jun} depending on the material of the ball in order to ensure a constant central film thickness of 100 nm required for calibration under the conditions of 0.583 GPa average contact pressure and 40 $\mathrm{^o}$C inlet oil temperature for all the tests. In addition to the solid temperatures, traction coefficients measured independently using MTM2 rig are used in the current validation. To accurately simulate the thermal cases, the Roelands and Eyring models (Eqs. \ref{eq:Exp_Roelands} and \ref{eq:Exp_Eyring} respectively) \cite{Hartinger2018Oct} are used to characterise the viscosity of Santotrac 50 due to pressure, temperature and shear-thinning effects, while the Yasutomi model coupled with the Carreau equation (Eqs. \ref{eq:Exp_Yasutomi} and \ref{eq:Exp_Carreau}, respectively) \cite{Bair2019Feb} are used to model the PAO 4 viscous behaviour. The density variation of both fluids are numerically describe using the Dowson and Higginson model \cite{Zhu2019Dec} while their thermal conductivity and heat capacity are correlated using the models by Larsson and Anderson \cite{Larsson2000Apr}. Readers may refer to \cite{Hartinger2018Oct, Bair2019Feb, Larsson2000Apr} for the input parameters of the corresponding relationships, while the solids thermal and mechanical properties can be found in \cite{Lu2020Jun}.
\begin{subequations}
    \begin{align}
        \eta &= \eta_0 \; \exp \bigg( ( \ln \eta_0 + 9.67 ) \times \bigg[ -1 + \left( 1 + 5.1 \times 10^{-9} p \right)^{\dfrac{\alpha}{5.1 \times 10^{-9} \left( \ln \eta_{0} + 9.67 \right)}} \left( \frac{T - 138}{T_0 - 138} \right)^{- \dfrac{\beta (T_{0} - 138)}{\left( \ln \eta_{0} + 9.67 \right)}} \bigg] \bigg),
        \label{eq:Exp_Roelands}
        \\[1ex]
        \eta^{*} &= \eta \left( \tau_{e} / \tau_{0} \right) \bigg/ \sinh{\left( \tau_{e} / \tau_{0} \right)},
        \label{eq:Exp_Eyring}
        \\[1ex]
        \mu &= \mu_{g} \exp \left[ \frac{-2.303 C_{1} \left( T - \left[T_{g0} + A_{1} \ln\left(1 + A_{2}p \right) \right] \right) \left( 1 + b_{1} p \right)^{b_2}}{C_{2} + \left( T - \left[T_{g0} + A_{1} \ln\left(1 + A_{2}p \right) \right] \right) \left( 1 + b_{1} p \right)^{b_2}} \right],
        \label{eq:Exp_Yasutomi}
        \\[1ex]
        \eta &= \mu \left[ 1 + \left( \frac{\tau_e}{18 \times 10^6} \right)^{2} \right]^{-6/13},
        \label{eq:Exp_Carreau}
        \\
        \rho &= \rho_{0} \left[ 1 + \frac{0.6 \times 10^{-9} p}{1 + 1.7 \times 10^{-9} p} \right] \left[ 1 - 0.00065 \left( T - T_0 \right) \right],
        \label{eq:Exp_Dowson}
        \\[1ex]
        k &= k_{0} \left( 1 + \frac{c_{1} p}{1 + c_{2} p} \right),
        \label{eq:Exp_Conductivity}
        \\[1ex]
        \rho c_{p} &= \left( \rho c_{p} \right)_{0} \left[ 1 + \beta(p) (T - T_{0}) \left( 1 + \frac{k_{1} p}{1 + k_{2} p} \right) \right].
        \label{eq:Exp_SpecificHeat}
    \end{align}
\end{subequations}

\begin{figure}[H]
\centering
   \includegraphics[width=\linewidth]{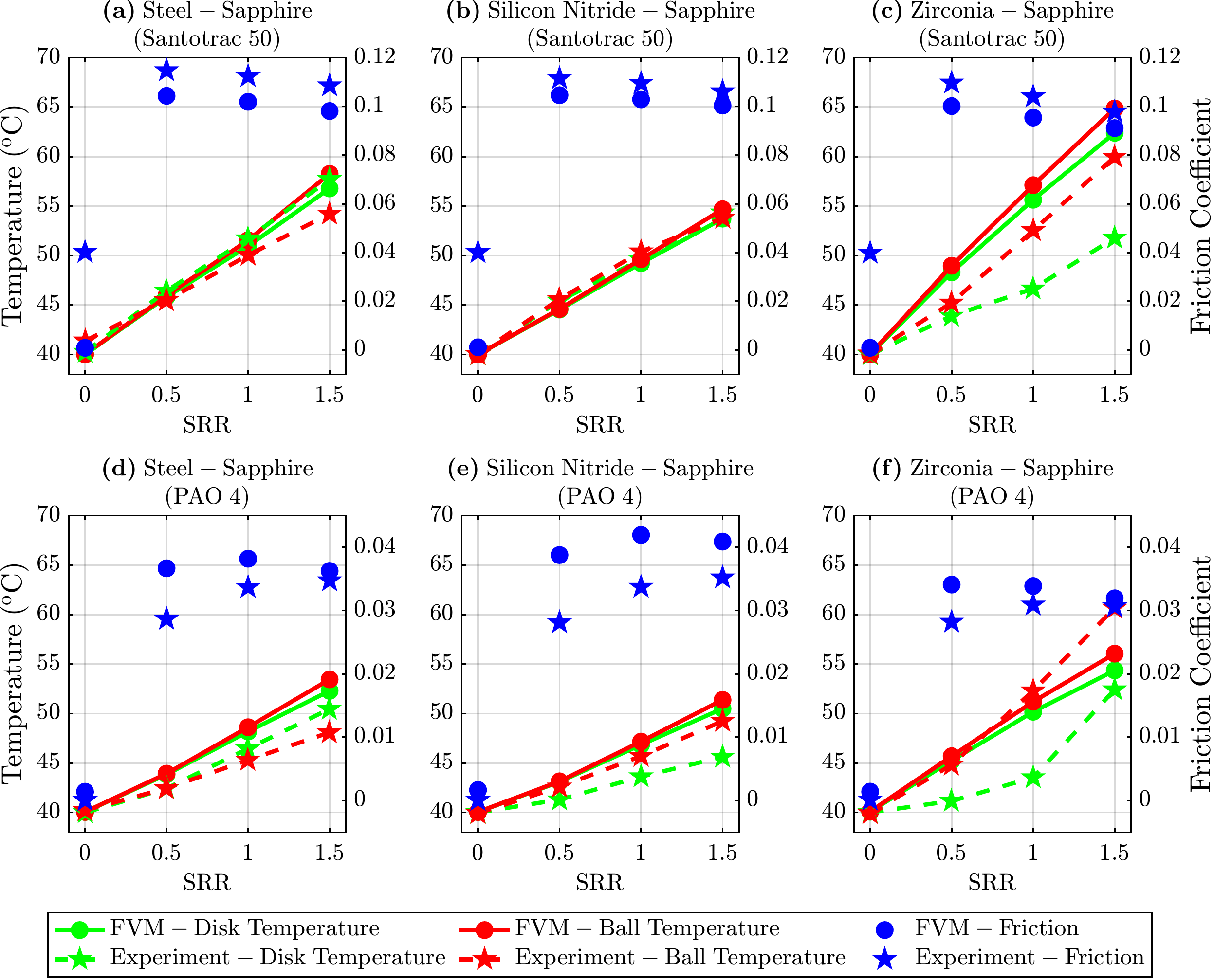}
   \caption{Average surface temperatures and traction coefficients for a ball on disk contact simulated using the finite volume framework (FVM) and validated using experimental results measured via infrared microscopy and MTM2 test rig for different ball materials and a sapphire disk; both solids bounding either Santotrac 50 or PAO 4 oil. The solid-fluid pairs and their respective entertainment speeds ($\mathrm{u_{e}}$) are: (\textbf{a}) steel-sapphire-Santotrac 50 ($\mathrm{u_{e}=0.308 \; m/s}$), (\textbf{b}) silicon nitride-sapphire-Santotrac 50 ($\mathrm{u_{e}=0.293 \; m/s}$), (\textbf{c}) zirconia-sapphire-Santotrac 50 ($\mathrm{u_{e}=0.308 \; m/s}$), (\textbf{d}) steel-sapphire-PAO 4 ($\mathrm{u_{e}=0.929 \; m/s}$), (\textbf{e}) silicon nitride-sapphire-PAO 4 ($\mathrm{u_{e}=0.806 \; m/s}$) and (\textbf{f}) zirconia-sapphire-PAO 4 ($\mathrm{u_{e}=0.929 \; m/s}$). All cases are carried out for SRR = [0, 0.5, 1.0, 1.5] and applied loads $\mathrm{W_{steel-sapphire}=20 \; N}$, $\mathrm{W_{zirconia-sapphire}=20 \; N}$ and $\mathrm{W_{steel-silicon \; nitride}=12 \; N}$, regardless of the entraining oil.}
   \label{fig:Experiment_1} 
\end{figure}

It is evident from Fig. \ref{fig:Experiment_1} that the current framework correctly predicts an increase in solid temperature with SRR due to the increase in heat generation. Vis-à-vis the contacts involving steel and silicon nitride with Santotrac 50 and PAO 4 (Figs. \ref{fig:Experiment_1}\textbf{a}, \textbf{b}, \textbf{d} and \textbf{e}), a considerable good match is obtained for the solid temperatures between those obtained numerically using the framework and their counterparts measured experimentally. However, the temperatures of the steel ball in Fig. \ref{fig:Experiment_1}\textbf{d} and sapphire disk in Fig. \ref{fig:Experiment_1}\textbf{e} at SRR = 1.5 are found to be numerically overestimated with a 10-11 \% deviation from their experimental results. This could be attributed to the film thickness and its impact on the proportion of generated heat in the contact; unlike the tests during which the central film thickness was kept constant at 100 nm, the developed framework computes a 110 nm central thickness for both cases as it is unfeasible to impose a constant thickness in the simulation due to the influence of pressure, temperature and shear-rate effects on the film geometry. On the other hand, it is obvious that significant discrepancies between the measured and computed average solid temperatures are found for the cases involving the zirconia ball, with either Santotrac 50 or PAO 4 (Figs, \ref{fig:Experiment_1}\textbf{c} and \textbf{f}, respectively). This is not surprising as zirconia is characterised as a highly effective thermal resistance material, thus possessing an ability to hinder the flow of heat through its depth, which consequently results in high temperatures at the outer layer of the solid where a significant proportion of heat is accumulated. As a result of the near-infrared radiative properties of zirconia, the utilised infrared camera in \cite{Lu2020Jun} was incapable of accurately measuring the temperature distributions which led the authors to implement Carslaw \& Jaeger’s equation, as reported in \cite{Lu2020Jun}.

Furthermore, traction coefficient data illustrated in Fig. \ref{fig:Experiment_1} confirm the validity of the developed numerical models in predicting the lubrication performance. The only noticeable discrepancy between experiments and numerical results is obtained under pure-rolling conditions in cases where Santotrac 50 is applied (Figs. \ref{fig:Experiment_1}\textbf{a}, \textbf{b} and \textbf{c}). This could be attributed to the limitations which could occur when controlling the thickness at a prescribed level in cases where surfaces travel at similar speeds, especially considering that a good match in friction coefficients is obtained at the same speed conditions but when PAO 4 is used (Figs. \ref{fig:Experiment_1}\textbf{d}, \textbf{e} and \textbf{f}). Moreover, the good match in friction coefficients obtained with the zirconia ball shows that the numerically predicted heat generation is accurately predicted, ergo elucidating the aforementioned reasoning to identify what may have caused the mismatch in measure and computed average temperatures.

\begin{figure}[H]
\centering
   \includegraphics[width=\linewidth]{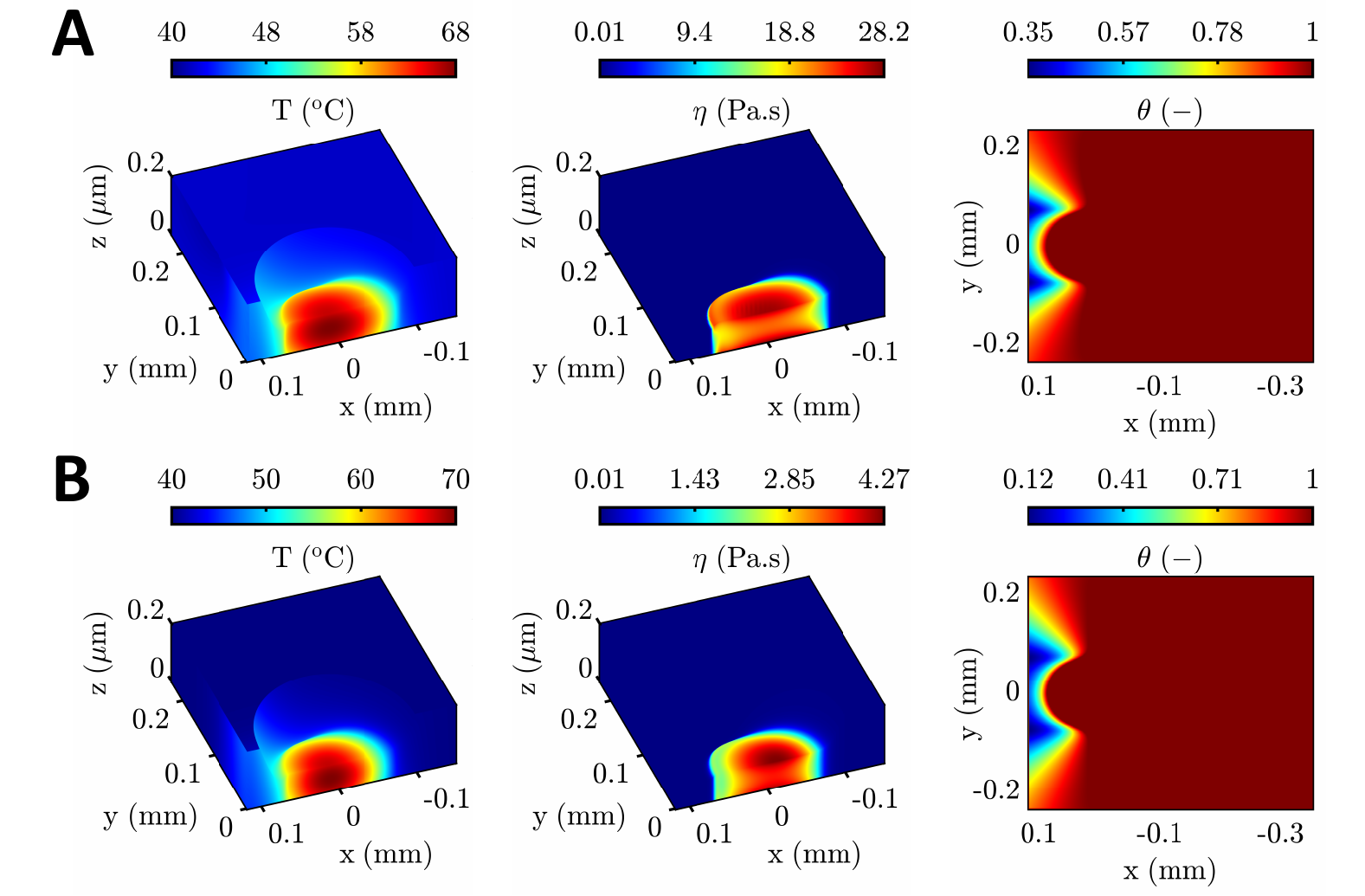}
   \caption{Three-dimensional cross-sectional view depicting the mid-plane distribution of the film temperature ($\mathrm{T}$), film dynamic viscosity ($\mathrm{\eta}$) and a two-dimensional illustration of the film content ($\mathrm{\theta}$) at slide-roll-ratio SRR = 1.5 for a silicon nitride ball loaded against a sapphire disk (W = 20 N), where the two solids are separated using \textbf{(A)} Santotrac 50 oil at entrainment speed $\mathrm{u_{e}=0.293 \; m/s}$ and \textbf{(B)} PAO 4 oil at entrainment speed $\mathrm{u_{e}=0.806 \; m/s}$. According to the complementary JFO conditions enforced by the Elrod-Adams cavitation algorithm, film rupture occurs when the fractional film content ($\mathrm{\theta}$) drops below unity, hence yielding in a biphasic mixture of gas/vapour, as shown near the contact outlet.}
   \label{fig:Experiment_2} 
\end{figure}

To achieve a clear visualisation of the FVM thermal results presented in Figs. \ref{fig:Experiment_1}, and particularly those related to the silicon nitride and sapphire contact case at SRR = 1.5, Fig. \ref{fig:Experiment_2} illustrates three-dimensional cross-sectional maps of the film temperature, dynamic viscosity as well as a two-dimensional representation of the fluid film fraction dictated by the solution of the $(p-\theta)$ generalised Reynolds equation, where Figs. \ref{fig:Experiment_2}\textbf{A} and \ref{fig:Experiment_2}\textbf{B} correspond to Santotrac 50 and PAO4 oils, respectively. Regarding the temperatures of the two films, it is evident that shear heating effects are more dominant in the case of PAO4, hence yielding in slightly higher temperatures compared to Santotrac 50. The direct influence of temperature on film viscosity ergo yields much lower PAO4 viscosity values compared to Santotrac 50, the latter being characterised by a more uniform viscosity distribution across the contact length. However, it is important to note that different rheological models were used to account for the behaviour of Santotrac 50 and PAO4 as previously described, due to the limited availability of relevant parameters; hence this can potentially play an important role in these comparisons. Film fraction maps, which are rarely discussed in the field of TEHL simulations, illustrate how film rupture commences near the outlet of the contact, particularly where the divergent zone is found, where a higher liquid-vapour/gases mixture can be noticed within the cavitation zone in the case of PAO4 in Fig. \ref{fig:Experiment_2}\textbf{B} (12\% liquid compared to 88\% mixture of liquid and gases/vapour).

\section{Conclusion}
\label{sec:Conclusion}
We have presented a finite volume fluid-structure interaction (FSI) framework for solving thermal-EHL problems mapped on non-orthogonal curvilinear grids using the strong conservation form of the pertinent governing equations. Numerical and experimental benchmarks demonstrated the accuracy of the developed TEHL simulation framework in predicting the performance of lubricated interfaces, where pressure, film thickness, friction forces and fluid and solid temperature profiles have been determined. The finite volume implementation facilitates the numerical discretisation of the governing equations in the computational domain, including the convective and diffusive cross-derivative terms of the fluid energy equation resulting from the non-orthogonal grid mapping, while maintaining the virtue of conservation property in the discrete level of finite volume methods, which is expensive to other discretisation methods. Furthermore, the Elrod-Adams $(p-\theta)$ cavitation model is combined with the generalised Reynolds equation to account for fluid film cavitation and detect the formation of biphasic mixtures within the contact. The effectiveness of the developed framework for coupling the TEHL solvers was demonstrated by simulating lubricated counterformal contacts under various operating conditions. The architecture of the framework and the additional implementation of the semi-system approach and Point Gauss-Seidel Method with Aitken Acceleration (PGMA) for the EHL solution revamped the numerical convergence and significantly decreased the computational cost of the simulations. 

The current contribution may be regarded as an enhancement to traditional Reynolds-based solvers due to the limitations encountered in considering any curvature effects when modelling FSI problems in the presence of complex geometries. Although commercial CFD codes may excel in tackling such problems due to their powerful resources, yet they suffer from higher computational overhead time and additional complexities in generating surface meshes. The proposed method of mapping the strong conservation form of the TEHL governing equations on a computational domain using Reynolds-based solvers overcomes these limitations, which will be further tested by analysing the details of fluid recirculation and temperature gradients at the contact inlet compared with both CFD and modified Reynolds solvers using standard Cartesian grids. This method holds promise and will be further investigated in applications where curvature effects and cross-derivative terms are more profound than counterformal contacts, such as journal bearing conformal systems, which the authors will study in future contributions.


\newpage
\appendix

\newpage
\section{Coordinate Transformation from Cartesian to Curvilinear Grid System}
\label{subsection:Coordinate_Transformation}
The time-dependent coordinate transformation of the physical Cartesian space $\chi^{3}_{t} \subset \mathbb{R}^{3}$ from the transformed generalised curvilinear space $\varepsilon^{3}_{\tau} \subset \mathbb{R}^{3}$, where $\mathbb{R}^{3}$ is the three-dimensional Euclidian space, can be expressed as \cite{Liseikin2017Jun, Farrashkhalvat2003Mar}
\begin{gather} \label{eq:Coord_Transformation}
     \boldsymbol{x \left( \xi,\tau \right)} : \varepsilon^{3}_{\tau} \rightarrow \chi^{3}_{t},
     \quad \quad
     \boldsymbol{\xi} = \left( \xi, \eta, \gamma, \tau \right) \in \varepsilon^{3}_{\tau},
     \quad \quad
     \boldsymbol{x} = \left( x, y, z, t \right) \in \chi^{3}_{t}, \nonumber
     \\[1ex]
     x^{i} = x^{i} \left( \xi^{j}, \tau \right) = x^{i} \left( \xi, \eta, \gamma, \tau \right),
     \quad
     \left(i,j = 1,2,3 \right),
\end{gather}
where $\boldsymbol{x \left( \xi, \tau \right)}$ is assumed to be smooth and invertible with respect to $\xi^{j}$ and $\tau$. Therefore, the time-dependent inverse transformation can be written as
\begin{gather}
     \boldsymbol{\xi \left( x, t \right)} : \chi^{3}_{t} \rightarrow \varepsilon^{3}_{\tau}, \nonumber
     \\[1ex]
     \xi^{i} = \xi^{i} \left( x^{j}, t \right) = \xi^{i} \left( x, y, z, t \right),
     \quad
     \left(i,j = 1,2,3 \right).
\end{gather}
In the above equations, $x^{i}$ are the Cartesian coordinates of a point in the physical space which at time $t$ has curvilinear coordinates $\xi^{j}$. The time-dependent coordinate transformation (Eq. \ref{eq:Coord_Transformation}) allows the solution of transport equations in time-dependent moving grids in physical space. 

The coordinate transformation $\boldsymbol{x \left( \xi, \tau \right)}$ is characterised by the transformation tensor $A^{i}_{j}$, also known as the covariant basis matrix, and the determinant known as Jacobian $J$ defined as
\begin{align} 
     A^{i}_{j} = \left( \frac{\partial x^{i}}{\partial \xi^{j}} \right)
     \quad
     \Rightarrow
     \quad
     \left[ \boldsymbol{A} \right] = 
       \begin{pmatrix}
         x_{\xi} & x_{\eta} & x_{\gamma} \\
         y_{\xi} & y_{\eta} & y_{\gamma} \\
         z_{\xi} & z_{\eta} & z_{\gamma} 
       \end{pmatrix},
     \quad
     J = det \left( \frac{\partial x^{i}}{\partial \xi^{j}} \right) = det \left[ \boldsymbol{A} \right],
     \label{eq:Coord_Transformation_Matrix}
\end{align}
where $x_{\xi}$ and $z_{\eta}$ are shorthand notations for the partial derivatives $\dfrac{\partial x}{\partial \zeta}$ and $\dfrac{\partial z}{\partial \eta}$, respectively. Similarly, the inverse coordinate transformation $\boldsymbol{\xi \left( x, t \right)}$, known as the contravariant basis matrix, is characterised by the inverse transformation tensor and Jacobian as follows
\begin{align} 
     \bar A^{i}_{j} = \left( \frac{\partial \xi^{i}}{\partial x^{j}} \right)
     \quad
     \Rightarrow
     \quad
     \left[ \boldsymbol{\bar{A}} \right] = 
       \begin{pmatrix}
         \xi_{x}    & \xi_{y}    & \xi_{z} \\
         \eta_{x}   & \eta_{y}   & \eta_{z} \\
         \gamma_{x} & \gamma_{y} & \gamma_{z} 
       \end{pmatrix},
     \quad
     \frac{1}{J} = det \left( \frac{\partial \xi^{i}}{\partial x^{j}} \right) = det \left[ \boldsymbol{\bar{A}} \right].
     \label{eq:Inverse_Coord_Transformation_Matrix}
\end{align}

The transformation tensors are correlated by the identity $A^{i}_{j} \bar{A}^{k}_{j} = \delta^{k}_{j}$, or in matrix array notation, $\boldsymbol{\left[ A \right] \left[ \bar{A} \right] = \left[ I \right]}$ $\Rightarrow$ $\boldsymbol{\left[ \bar{A} \right] = \left[ A \right]^{-1}}$. Therefore, the relationship between the elements of $\boldsymbol{\left[ A \right]}$ and $\boldsymbol{\left[ \bar{A} \right]}$ can be written as 
\begin{subequations}
    \begin{align}
     \frac{\partial \xi^{i}}{\partial x^{j}} 
     &= 
     \frac{1}{J} \left( \frac{\partial x^{j+1}}{\partial \xi^{i+1}} \frac{\partial x^{j+2}}{\partial \xi^{i+2}} - \frac{\partial x^{j+1}}{\partial \xi^{i+2}} \frac{\partial x^{j+2}}{\partial \xi^{i+1}} \right), 
     \\[1ex]
     \frac{\partial x^{i}}{\partial \xi^{j}} 
     &= 
     J \left( \frac{\partial \xi^{j+1}}{\partial x^{i+1}} \frac{\partial \xi^{j+2}}{\partial x^{i+2}} - \frac{\partial \xi^{j+1}}{\partial x^{i+2}} \frac{\partial \xi^{j+2}}{\partial x^{i+1}} \right).
    \end{align}
\end{subequations}
In the above equations, the superscript indices follow the cyclic permutation order (1,2,3). With this convention, any index, say $l$, is identified with $l \pm 3$; thus, for instance, $\dfrac{\partial x^{4}}{\partial \xi^{5}} = \dfrac{\partial x^{1}}{\partial \xi^{2}} = \dfrac{\partial x}{\partial \eta}$.

The covariant base vectors $\boldsymbol{g_{i}}$ which describe how a point in space changes as the coordinates change can be expressed as
\begin{align} 
     \boldsymbol{g_{i}} = \left( \frac{\partial x^{j}}{\partial \xi^{i}} \right) \boldsymbol{i_{j}} = A^{j}_{i} \boldsymbol{i_{j}} 
     \Rightarrow
     \begin{bmatrix}
       \boldsymbol{g_{1}} \\
       \boldsymbol{g_{2}} \\
       \boldsymbol{g_{3}} 
     \end{bmatrix}
     =
     \left[ \boldsymbol{A} \right]^{T}
     \begin{bmatrix}
       \boldsymbol{i_{1}} \\
       \boldsymbol{i_{2}} \\
       \boldsymbol{i_{3}} 
     \end{bmatrix},
     \label{eq:Covariant_BaseVector}
\end{align}
while the contravariant base vectors $\boldsymbol{g^{i}}$ that describe how the coordinates change as the point in space changes can be written as
\begin{align} 
     \boldsymbol{g^{i}} = \left( \frac{\partial \xi^{i}}{\partial x^{j}} \right) \boldsymbol{i_{j}} = A^{j}_{i} \boldsymbol{i_{j}} 
     \Rightarrow
     \begin{bmatrix}
       \boldsymbol{g^{1}} \\
       \boldsymbol{g^{2}} \\
       \boldsymbol{g^{3}} 
     \end{bmatrix}
     =
     \left[ \boldsymbol{\bar{A}} \right]
     \begin{bmatrix}
       \boldsymbol{i_{1}} \\
       \boldsymbol{i_{2}} \\
       \boldsymbol{i_{3}} 
     \end{bmatrix}.
     \label{eq:Contravariant_BaseVector}
\end{align}
The covariant and contravariant metric coefficients, $g_{ij}$ and $g^{ij}$ respectively, can be calculated by the scalar products of the covariant and contravariant bases in general curvilinear coordinates with non-orthonormal bases \cite{Nguyen-Schafer} as follows
\begin{subequations}  \label{eq:Metric_Coeffs}
    \begin{align}
     g_{ij} & = \boldsymbol{g}_{i} \cdot \boldsymbol{g}_{j} = \frac{\partial x^{k}}{\partial \xi^{i}} \frac{\partial x^{k}}{\partial \xi^{j}}
     \Rightarrow
     \boldsymbol{\left[ G \right]}  = \boldsymbol{\left[ A \right]^{T}} \boldsymbol{\left[ A \right]} = \boldsymbol{\left[ G \right]^{T}},
     \\[1ex]
     g^{ij} & = \boldsymbol{g}^{i} \cdot \boldsymbol{g}^{j} = \frac{\partial \xi^{i}}{\partial x^{k}} \frac{\partial \xi^{j}}{\partial x^{k}}
     \Rightarrow
     \boldsymbol{\left[ \bar{G} \right]}  = \boldsymbol{\left[ \bar{A} \right]} \boldsymbol{\left[ \bar{A} \right]^{T}} = \boldsymbol{\left[ \bar{G} \right]^{T}},
    \end{align}
\end{subequations}
in addition to the physical covariant base vectors which can be evaluated using
\begin{align} 
     \boldsymbol{e_{i}} = \left( 1/\sqrt{g_{ii}} \right) \boldsymbol{g_{i}}.
\end{align}

The metric coefficients expressed in Eq. \ref{eq:Metric_Coeffs} can be employed to alternate between the covariant and contravariant components of a vector. Suppose a vector field $\boldsymbol{\upsilon}$ is a function of position vector $\boldsymbol{r}$ at an arbitrary point P in space, $\boldsymbol{\upsilon}$ can be expressed in terms of components in the: global Cartesian basis $\{ \boldsymbol{i}_i \}$, local contravariant basis $\{ \boldsymbol{g}^i \}$, local covariant basis $\{ \boldsymbol{g}_i \}$, or physical covariant basis $\{ \boldsymbol{e}_i \}$. Using the Einstein summation notation, this implies
\begin{align}
 \begin{split} \label{eq:Vector_Notation}
     \boldsymbol{\upsilon} & = u_{i} \boldsymbol{i}_{i} = \upsilon_{i} \boldsymbol{g}^{i} = \upsilon^{i} \boldsymbol{g}_{i} = \hat{\upsilon}^{i} \boldsymbol{e}_{i} ,
     \\
     \boldsymbol{\upsilon} & = u \boldsymbol{i} + v \boldsymbol{j} + w \boldsymbol{k}
     \\
     & = v_{1} \boldsymbol{g}^{1} + v_{2} \boldsymbol{g}^{2} + v_{3} \boldsymbol{g}^{3}
     \\
     & = v^{1} \boldsymbol{g}_{1} + v^{2} \boldsymbol{g}_{2} + v^{3} \boldsymbol{g}_{3}
     \\
     & = \hat{v}^{1} \boldsymbol{e}_{1} + \hat{v}^{2} \boldsymbol{e}_{2} + \hat{v}^{3} \boldsymbol{e}_{3},
 \end{split}
\end{align}
where $u_{i}$, $v_{i}$, $v^{i}$ and $\hat{v}^{i}$ are the Cartesian, covariant, contravariant, and physical components of $\boldsymbol{\upsilon}$, respectively. The relationships between these components can be written as
\begin{align*}
     v_{i} = \left( \frac{\partial x^{j}}{\partial \xi^{i}} \right) u_{j} = A^{j}_{i} u_{j} = g_{ij} v^{j} \Rightarrow 
     \begin{bmatrix}
       v_{1} \\
       v_{2} \\
       v_{3} 
     \end{bmatrix}
     =
     \left[ \boldsymbol{A} \right]^{T}
     \begin{bmatrix}
       u \\
       v \\
       w 
     \end{bmatrix}
     =
     \left[ \boldsymbol{G} \right]
     \begin{bmatrix}
       v^{1} \\
       v^{2} \\
       v^{3} 
     \end{bmatrix},
\end{align*}

\begin{align*}
     v^{i} = \left( \frac{\partial \xi^{i}}{\partial x^{j}} \right) u_{j} = \bar{A}^{i}_{j} u_{j} = g^{ij} v_{j} \Rightarrow 
     \begin{bmatrix}
       v^{1} \\
       v^{2} \\
       v^{3} 
     \end{bmatrix}
     =
     \left[ \boldsymbol{\bar{A}} \right]
     \begin{bmatrix}
       u \\
       v \\
       w 
     \end{bmatrix}
     =
     \left[ \boldsymbol{\bar{G}} \right]
     \begin{bmatrix}
       v_{1} \\
       v_{2} \\
       v_{3} 
     \end{bmatrix},
\end{align*}
\begin{align*}
     \hat{v}^{i} = \sqrt{g_{ii}} v^{i} \Rightarrow 
     \begin{bmatrix}
       \hat{v}^{1} &
       \hat{v}^{2} &
       \hat{v}^{3} 
     \end{bmatrix}^{T}
     =
     \sqrt{diag \left[ \boldsymbol{G} \right]}
     \begin{bmatrix}
       v^{1} &
       v^{2} &
       v^{3} 
     \end{bmatrix}^{T},
\end{align*}
\begin{align*}
 \begin{split}
    u_{i} & = \left( \frac{\partial \xi^{j}}{\partial x^{i}} \right) v_{j} = \bar{A}^{j}_{i} v_{j} \Rightarrow 
     \begin{bmatrix}
       u &
       v &
       w 
     \end{bmatrix}^{T}
     =
     \left[ \boldsymbol{\bar{A}} \right]^{T}
     \begin{bmatrix}
       v_{1} &
       v_{2} &
       v_{3} 
     \end{bmatrix}^{T}
    \\
    & = \left( \frac{\partial x^{i}}{\partial \xi^{j}} \right) v^{j} = A^{i}_{j} v^{j} \Rightarrow 
     \begin{bmatrix}
       u &
       v &
       w 
     \end{bmatrix}^{T}
     =
     \left[ \boldsymbol{A} \right]^{T}
     \begin{bmatrix}
       v^{1} &
       v^{2} &
       v^{3} 
     \end{bmatrix}^{T}
    \\
    & = \left( \frac{\partial x^{i}}{\partial \xi^{j}} \right) \frac{\hat{v}^{j}}{\sqrt{g_{ii}}} = \left( \frac{A^{i}_{j}}{\sqrt{g_{jj}}} \right) \hat{v}^j.
 \end{split}
\end{align*}

To represent the conservation equations in the strong conservation form using generalised coordinates, it is convenient to express the flow velocity field with the Cartesian components $u_{i} \left(u,v,w\right)$ and use the contravariant components $v^{i} \left(v^{1},v^{2},v^{3}\right)$ for the advection velocity \cite{Ogawa1987, Yang1994, Vinokur1974Feb}. Therefore, it is only necessary to transform the spatial derivatives with respect to Cartesian coordinates into generalised coordinates. For a general scalar field $\varphi$, this transformation can be written as

\begin{align}
\label{eq:CoordTrans:SpatialDerivative}
    \frac{\partial \varphi}{\partial x^{i}} = \frac{\partial \xi^{j}}{\partial x^{i}} \frac{\partial \varphi}{\partial \xi^{j}}
    =
    \bar{A}^{j}_{i} \frac{\partial \varphi}{\partial \xi^{j}}
    \Rightarrow
    \begin{bmatrix}
      \dfrac{\partial \varphi}{\partial x} & \dfrac{\partial \varphi}{\partial y} & \dfrac{\partial \varphi}{\partial z}
    \end{bmatrix}^{T}
    =
    \left[ \boldsymbol{\bar{A}} \right]^{T}
    \begin{bmatrix}
      \dfrac{\partial \varphi}{\partial \xi} & \dfrac{\partial \varphi}{\partial \eta} & \dfrac{\partial \varphi}{\partial \gamma}
    \end{bmatrix}^{T}.
\end{align}

\newpage
\section{General Strong Scalar Conservation Equation}
\label{subsection:General_Strong_Conservation_Equation}
The general conservation equation for an arbitrary scalar physical quantity $\phi$ can be written in coordinate-free vector form as \cite{Ferziger2020}
\begin{align} 
    \frac{\partial \left( \rho \phi \right)}{\partial t} + \boldsymbol{\nabla} \cdot \left( \rho \boldsymbol{\nu} \phi \right) = \boldsymbol{\nabla} \cdot \left( \Gamma \boldsymbol{\nabla} \phi \right) + S_{\phi},
\end{align}
where $\rho$ is the fluid density, $\boldsymbol{\nu}$ is the velocity vector field of the fluid, $\Gamma$ is the diffusion coefficient, and $S_{\phi}$ denotes the source/sink term. This equation is often referred to as a strong conservation (or divergent) form of the governing transport equation of $\phi$. In Cartesian coordinates $x_{i}(x,y,z)$, this equation is expressed in tensor notation as \cite{Ferziger2020}
\begin{align} \label{eq:Cartesian_Phi}
    \frac{\partial \left( \rho \phi \right)}{\partial t} + \frac{\partial}{\partial x^{j}} \left(  \rho u_{j} \phi \right) = \frac{\partial}{\partial x^{j}} \left( \Gamma \frac{\partial \phi}{\partial x^{j}} \right) + S_{\phi},
\end{align}
where $u_{i}(u,v,w)$ are the Cartesian components of the velocity vector, i.e., $\boldsymbol{\nu} = u\boldsymbol{i} + v\boldsymbol{j} + w\boldsymbol{k}$, $ \{ \boldsymbol{i},\boldsymbol{j},\boldsymbol{k} \}$ being the unit vectors in the direction of the Cartesian axes (Cartesian basis).

Considering a generalised curvilinear coordinate system defined by the time-independent coordinate transformation given in Eq. \ref{eq:Coord_Transformation}, the general scalar conservation equation may be written in the contravariant components form as \cite{Ogawa1987, Kajishima, Yang1994, Vinokur1974Feb, Liseikin2017Jun, Daiguji1993}
\begin{align} \label{eq:Curvilinear_Phi}
    \frac{\partial \left( J \rho \phi \right)}{\partial t} + \frac{\partial}{\partial \xi^{j}} \left( J \rho \nu^{j} \phi \right) = \frac{\partial}{\partial \xi^{j}} \left[ J g^{kj} \Gamma \left( \frac{\partial \phi}{\partial \xi^{k}} \right) \right] + JS_{\phi}, \quad \quad \quad g^{kj} = \frac{\partial \xi^{k}}{\partial x^{m}} \frac{\partial \xi^{j}}{\partial x^{m}},
\end{align}
where $\nu^{j}$ are the contravariant components of the flow velocity vector in the local basis of the tangential vectors $\{ \boldsymbol{g_{j}} \}$ (covariant basis). These components are proportional to the physical components of the velocity vector tangential to the coordinate curve $\xi^{j}$ at any point in space.

\newpage
\section{Mass Conservation Equation (Continuity Equation)}
\label{subsection:Continuity_Equation}
The mass conservation equation for compressible fluids is expressed in coordinate-free vector form as \cite{White2011Mar}
\begin{align}
    \frac{D \rho}{D t} + \rho \boldsymbol{\nabla} \cdot \boldsymbol{v} = 0 \Leftrightarrow \frac{\partial \rho}{\partial t} + \boldsymbol{\nabla} \cdot \left( \rho \boldsymbol{v} \right) = 0,
\end{align}
where $\rho$ is the fluid density and $\boldsymbol{v}$ the velocity vector field of the fluid. This implies that the continuity equation in Cartesian coordinates can be written in tensor notation as
\begin{align}
    \frac{D \rho}{D t} + \rho \frac{\partial u_{i}}{\partial x_{i}} = 0 \Leftrightarrow \frac{\partial \rho}{\partial t} + \frac{\partial \left( \rho u_{i} \right)}{\partial x_{i}} = 0.
\end{align}

Therefore, considering the coordinates transformation described in Appendix \ref{subsection:Coordinate_Transformation} and the normalisation of the dependent variables summarized in Appenidx \ref{sec:Appendix_D}, the dimensionless continuity equation derived in the generalised transformed domain can be expressed in tensor notation as
\begin{align}
    \frac{\partial \Bar{\rho}}{\partial \tau} + \nabla_{X} \cdot \left( \Bar{\rho} \boldsymbol{\Tilde{V}} \right) = 0,
\end{align}
where the components of $\boldsymbol{\Tilde{V}} = \left[ \Tilde{V}^X \; \Tilde{V}^Y \; \Tilde{V}^Z \right]$ are evaluated using Eqs. \ref{eq:Ux}, \ref{eq:Vx} and \ref{eq:Wx}.

\newpage
\section{Integral Terms of the Generalised Reynolds Equation and Velocity Components}
\label{sec:Appendix_A}
The integral terms of the p-$\theta$ modified generalised Reynolds equation and velocity components expressed using Cartesian coordinates are defined as
\begin{subequations}
  \begin{align}
    & \frac{1}{\eta_e} = \bigintssss_{z_1}^{z_2} \frac{dz}{\eta}, 
  \end{align}
  \begin{align}
    & \frac{1}{\eta_{e}^{'}} = \bigintssss_{z_1}^{z_2} \frac{z \; dz}{\eta},
  \end{align}
  \begin{align}
    & \rho^{'} =  \bigintssss_{z_1}^{z_2} \rho \; \left( \bigintssss_{z_1}^{z} \frac{dz'}{\eta} \right) \; dz,
  \end{align}
  \begin{align}
    & \rho^{''} =  \bigintssss_{z_1}^{z_2} \rho \; \left( \bigintssss_{z_1}^{z} \frac{z'dz'}{\eta} \right) \; dz,
  \end{align}
  \begin{align}
    & I(x,y,z) = \bigintssss_{z_1}^{z} \frac{z' \; dz'}{\eta},
  \end{align}
  \begin{align}
    & J(x,y,z) = \bigintssss_{z_1}^{z} \frac{dz'}{\eta}.
  \end{align}
  \begin{align}
    & I_{p} (x,y,z) = \left( I - J \frac{\eta_e}{\eta_{e}^{'}} \right),
  \end{align}
  \begin{align}
    & I_{s} (x,y,z) = J \eta_{e}.
  \end{align}
\end{subequations}

On the other hand, the coefficients and integral terms of p-$\theta$ mass-conserving generalised Reynolds equation and velocity components expressed using normalised transformed coordinates are given as
\begin{subequations}
  \begin{align}
    & \Bar{I}(X,Y,Z) = \bigintssss_{0}^{Z} \frac{Z' \; dZ'}{\Bar{\eta}},
  \end{align}
  \begin{align}
    & \Bar{J}(X,Y,Z) = \bigintssss_{0}^{Z} \frac{dZ'}{\Bar{\eta}},
  \end{align}
  \begin{align}
    & \Bar{I}_{1}(X,Y) = \Bar{I}(X,Y,1) = \bigintssss_{0}^{1} \frac{Z \; dZ}{\Bar{\eta}},
  \end{align}
  \begin{align}
    & \Bar{J}_{1}(X,Y) = \Bar{J}(X,Y,1) = \bigintssss_{0}^{1} \frac{dZ}{\Bar{\eta}},
  \end{align}
  \begin{align}
    & \Bar{I}_{p} (X,Y,Z) = H^{2} \left[ \Bar{I}(X,Y,Z) - \Bar{J}(X,Y,Z) \frac{\Bar{I}_{1}(X,Y)}{\Bar{J}_{1}(X,Y)} \right],
  \end{align}
  \begin{align}
    & \Bar{I}_{s} (X,Y,Z) = \frac{\Bar{J}(X,Y,Z)}{\Bar{J}_{1}(X,Y)},
  \end{align}
  \begin{align}
    & \Bar{R}_{1} = \bigintssss_{0}^{1} \Bar{\rho} \; \left( \bigintssss_{0}^{Z} \frac{dZ'}{\Bar{\eta}} \right) \; dZ,
  \end{align}
  \begin{align}
    & \Bar{R}_{2} = \bigintssss_{0}^{1} \Bar{\rho} \; \left( \bigintssss_{0}^{Z} \frac{Z' \; dZ'}{\Bar{\eta}} \right) \; dZ,
  \end{align}
  \begin{align}
    & \Bar{R}_{3} = \bigintssss_{0}^{1} \Bar{\rho} \; dZ,
  \end{align}
  \begin{align}
    & \Bar{\varepsilon} (X,Y)  = H^{3} \left[ \frac{\Bar{I}_{1}(X,Y) \Bar{R}_{1}(X,Y)}{\Bar{J}_{1}(X,Y)} - \Bar{R}_{2}(X,Y) \right],
  \end{align}
  \begin{align}
    & \Bar{\rho}^{*}_{e} (X,Y) = \frac{2 \Bar{R}_{1} (X,Y)}{\Bar{J}_{1} (X,Y)},
  \end{align}
  \begin{align}
    & \Bar{\rho}^{*}_{1} (X,Y) = \left[ \Bar{R}_{3} (X,Y) - \frac{2 \Bar{R}_{1} (X,Y)}{\Bar{J}_{1} (X,Y)} \right],
  \end{align}
  \begin{align}
    & \Bar{\rho}_{e} (X,Y) = \Bar{R}_{3} (X,Y).
  \end{align}
\end{subequations}

\newpage
\section{Coefficients of the Discrete $(p-\theta)$ Generalised Reynolds Equation in the Normalised Transformed Domain}
\label{sec:Appendix_B}
Assuming the variables computed at the CV faces to be equivalent to the average of the pertinent variables at the surrounding nodes, \emph{e.g.} $\left( \Bar{\varepsilon} \right)_{w} = \biggl[ \left( \Bar{\varepsilon} \right)_{W} + \left( \Bar{\varepsilon} \right)_{P} \biggr] \; \bigg/ \; 2 =  \biggl[ \left( \Bar{\varepsilon} \right)_{k, i-1} + \left( \Bar{\varepsilon} \right)_{k,i} \biggr] \; \bigg/ \; 2$, hence the coefficients in Eq. \ref{eq:Discrete_GRE} are defined as 
\begin{subequations}
    \begin{flalign}
        \alpha_{D} &=  \bigg[ \left( \Bar{\varepsilon} \right)_{w} \cdot \left( \delta Y \right)_{w} \big/ \left( \delta X \right)_{w} \bigg]
        \\
        \alpha_{C} &= - \left( \delta Y \right)_{w} \cdot \bigg[ \left( \Bar{\rho}^{*}_{e} U_{m} + \Bar{\rho}^{*}_{1} U_{1} \right)_{i} \cdot \theta_{i} \cdot D^{i,j}_{k-1,l} - \left( \Bar{\rho}^{*}_{e} U_{m} + \Bar{\rho}^{*}_{1} U_{1} \right)_{i-1} \cdot \theta_{i-1} \cdot D^{i-1,j}_{k-1,l} \bigg]
        \\
       \alpha &= \alpha_{D} + \alpha_{C} &&
    \end{flalign}
\end{subequations}
\begin{subequations}
    \begin{flalign}
        \begin{split}
           \beta_{D} &=  - \bigg[ \left( \Bar{\varepsilon} \right)_{w} \cdot \left( \delta Y \right)_{w} \big/ \left( \delta X \right)_{w} \bigg] - \bigg[ \left( \Bar{\varepsilon} \right)_{e} \cdot \left( \delta Y \right)_{e} \big/ \left( \delta X \right)_{e} \bigg] 
           \\
           &- r_{xy}^{2} \bigg[ \left( \Bar{\varepsilon} \right)_{n} \cdot \left( \delta X \right)_{n} \big/ \left( \delta Y \right)_{n} \bigg] - r_{xy}^{2} \bigg[ \left( \Bar{\varepsilon} \right)_{s} \cdot \left( \delta X \right)_{s} \big/ \left( \delta Y \right)_{s} \bigg]
        \end{split}
        \\
        \beta_{C} &= - \left( \delta Y \right)_{w} \cdot \bigg[ \left( \Bar{\rho}^{*}_{e} U_{m} + \Bar{\rho}^{*}_{1} U_{1} \right)_{i} \cdot \theta_{i} \cdot D^{i,j}_{k,l} - \left( \Bar{\rho}^{*}_{e} U_{m} + \Bar{\rho}^{*}_{1} U_{1} \right)_{i-1} \cdot \theta_{i-1} \cdot D^{i-1,j}_{k,l} \bigg]
        \\
       \beta &= \beta_{D} + \beta_{C} &&
    \end{flalign}
\end{subequations}
\begin{subequations}
    \begin{flalign}
        \gamma_{D} &=  \bigg[ \left( \Bar{\varepsilon} \right)_{e} \cdot \left( \delta Y \right)_{e} \big/ \left( \delta X \right)_{e} \bigg]
        \\
        \gamma_{C} &= - \left( \delta Y \right)_{e} \cdot \bigg[ \left( \Bar{\rho}^{*}_{e} U_{m} + \Bar{\rho}^{*}_{1} U_{1} \right)_{i} \cdot \theta_{i} \cdot D^{i,j}_{k+1,l} - \left( \Bar{\rho}^{*}_{e} U_{m} + \Bar{\rho}^{*}_{1} U_{1} \right)_{i-1} \cdot \theta_{i-1} \cdot D^{i-1,j}_{k+1,l} \bigg]
        \\
       \gamma &= \gamma_{D} + \gamma_{C} &&
    \end{flalign}
\end{subequations}
\begin{subequations}
    \begin{flalign}
        \zeta_{D} &= 0
        \\
        \zeta_{C} &= 0
        \\
        \zeta &= \zeta_{D} + \zeta_{C} &&
    \end{flalign}
\end{subequations}
\begin{subequations}
    \begin{flalign}
        \lambda_{D} &= 0
        \\
        \lambda_{C} &= 0
        \\
        \lambda &= \lambda_{D} + \lambda_{C} &&
    \end{flalign}
\end{subequations}
\begin{subequations}
    \begin{flalign}
        \varphi_{D} &= \left( \Bar{p} \right)_N \cdot r^{2}_{xy} \cdot \bigg[ \left( \Bar{\varepsilon} \right)_{n} \cdot \left( \delta X \right)_{n} \big/ \left( \delta Y \right)_{n} \bigg] + \left( \Bar{p} \right)_S \cdot r^{2}_{xy} \cdot \bigg[ \left( \Bar{\varepsilon} \right)_{n} \cdot \left( \delta X \right)_{s} \big/ \left( \delta Y \right)_{s} \bigg]
        \\
        \begin{split}
           \varphi_{C} &= \left( \Bar{\rho}^{*}_{e} U_{m} + \Bar{\rho}^{*}_{1} U_{1} \right)_{i} \cdot \theta_{i} \cdot \Bigg[ H_{i} - \bigg( \big[D^{i,j}_{k-1,l} \cdot \Bar{p}_{k-1,l} \big] - \big[ D^{i,j}_{k,l} \cdot \Bar{p}_{k,l} \big] - \big[ D^{i,j}_{k+1,l} \cdot \Bar{p}_{k+1,l} \big] \bigg) \Bigg]
           \\
           &- \left( \Bar{\rho}^{*}_{e} U_{m} + \Bar{\rho}^{*}_{1} U_{1} \right)_{i-1} \cdot \theta_{i-1} \cdot \Bigg[ H_{i-1} - \bigg( \big[D^{i-1,j}_{k-1,l} \cdot \Bar{p}_{k-1,l} \big] - \big[ D^{i-1,j}_{k,l} \cdot \Bar{p}_{k,l} \big] - \big[ D^{i-1,j}_{k+1,l} \cdot \Bar{p}_{k+1,l} \big] \bigg) \Bigg]
        \end{split}
        \\
       \varphi &= \varphi_{D} + \varphi_{C} &&
    \end{flalign}
\end{subequations}

\newpage
\section{Coefficients of the Discrete Energy Equation in the Normalised Transformed Domain}
\label{sec:Appendix_C}
Using the $\big[ \big[ \; \big] \big]$ operator of Pantankar where $\big[ \big[ A,B \big] \big]$ denotes the greater of $A$ and $B$ \cite{Patankar1980Jan}, and assuming the variables computed at the CV faces to be equivalent to the average of the pertinent variables at the surrounding nodes, \emph{e.g.} $\left( \Bar{k} \right)_{w} = \biggl[ \left( \Bar{k} \right)_{W} + \left( \Bar{k} \right)_{P} \biggr] \; \bigg/ \; 2 =  \biggl[ \left( \Bar{k} \right)_{k, i-1} + \left( \Bar{k} \right)_{k,i} \biggr] \; \bigg/ \; 2$, hence the coefficients in Eq. \ref{eq:Discrete_Energy} are defined as
\begin{subequations}
    \begin{flalign}
        \alpha_{D} &= -\left( \varepsilon \right)^{2} \cdot \Bar{k}_{w} \cdot H_{w} \cdot \mathcal{S}_{w} \big/ \left( \delta X \right)_{w}
        \\
        \alpha_{C} &= - \; \big[ \big[ \left( \Bar{\rho} \Bar{c}_{p} \Tilde{V}^{X} \right)_{w} \cdot \mathrm{Pe} \cdot \mathcal{S}_{w}, \; 0 \big] \big]
        \\
        \alpha &= \alpha_{D} + \alpha_{C} &&
    \end{flalign}
\end{subequations}
\begin{subequations}
    \begin{flalign}
        \gamma_{D} &= -\left( \varepsilon \right)^{2} \cdot \Bar{k}_{e} \cdot H_{e} \cdot \mathcal{S}_{e} \big/ \left( \delta X \right)_{e}
        \\
        \gamma_{C} &= - \; \big[ \big[ - \left( \Bar{\rho} \Bar{c}_{p} \Tilde{V}^{X} \right)_{e} \cdot \mathrm{Pe} \cdot \mathcal{S}_{e}, \; 0 \big] \big]
        \\
        \gamma &= \gamma_{D} + \gamma_{C} &&
    \end{flalign}
\end{subequations}
\begin{subequations}
    \begin{flalign}
        \zeta_{D} &= -\left( r_{xy} \varepsilon \right)^{2} \cdot \Bar{k}_{n} \cdot H_{n} \cdot \mathcal{S}_{n} \big/ \left( \delta Y \right)_{n}
        \\
        \zeta_{C} &= - \; \big[ \big[ - \left( \Bar{\rho} \Bar{c}_{p} \Tilde{V}^{Y} \right)_{n} \cdot \mathrm{Pe} \cdot \mathcal{S}_{n}, \; 0 \big] \big]
        \\
        \zeta &= \zeta_{D} + \zeta_{C} &&
    \end{flalign}
\end{subequations}
\begin{subequations}
    \begin{flalign}
        \lambda_{D} &= -\left( r_{xy} \varepsilon \right)^{2} \cdot \Bar{k}_{s} \cdot H_{s} \cdot \mathcal{S}_{s} \big/ \left( \delta Y \right)_{s}
        \\
        \lambda_{C} &= - \; \big[ \big[ \left( \Bar{\rho} \Bar{c}_{p} \Tilde{V}^{Y} \right)_{s} \cdot \mathrm{Pe} \cdot \mathcal{S}_{s}, \; 0 \big] \big]
        \\
        \lambda &= \lambda_{D} + \lambda_{C} &&
    \end{flalign}
\end{subequations}
\begin{subequations}
    \begin{flalign}
        \phi_{D} &= - \left( \Bar{k}_{t} \cdot \mathcal{S}_{t} \right) \cdot \bigg[ 1 
        + \left( \varepsilon^{2} \cdot \left[ \left( \nabla_{X} \cdot H \right)_{t} \right]^{2} \cdot Z_{t}^{2} \right) 
        + \left( \left( \varepsilon r_{xy} \right)^{2} \cdot \left[ \left( \nabla_{Y} \cdot H \right)_{t} \right]^{2} \cdot Z_{t}^{2} \right) 
        \bigg] \bigg/ \left( H \cdot \left( \delta Z \right)_{t} \right)
        \\
        \phi_{C} &= - \; \big[ \big[ - \left( \Bar{\rho} \Bar{c}_{p} \Tilde{V}^{Z} \right)_{t} \cdot \mathrm{Pe} \cdot \mathcal{S}_{t}, \; 0 \big] \big]
        \\
        \phi &= \phi_{D} + \phi_{C} &&
    \end{flalign}
\end{subequations}
\begin{subequations}
    \begin{flalign}
        \psi_{D} &= - \left( \Bar{k}_{b} \cdot \mathcal{S}_{b} \right) \cdot \bigg[ 1 
        + \left( \varepsilon^{2} \cdot \left[ \left( \nabla_{X} \cdot H \right)_{b} \right]^{2} \cdot Z_{b}^{2} \right) 
        + \left( \left( \varepsilon r_{xy} \right)^{2} \cdot \left[ \left( \nabla_{Y} \cdot H \right)_{b} \right]^{2} \cdot Z_{b}^{2} \right) 
        \bigg] \bigg/ \left( H \cdot \left( \delta Z \right)_{b} \right)
        \\
        \psi_{C} &= - \; \big[ \big[ \left( \Bar{\rho} \Bar{c}_{p} \Tilde{V}^{Z} \right)_{b} \cdot \mathrm{Pe} \cdot \mathcal{S}_{b}, \; 0 \big] \big]
        \\
        \psi &= \psi_{D} + \psi_{C} &&
    \end{flalign}
\end{subequations}
\begin{subequations}
    \begin{flalign}
        \beta_{D} &= - \alpha_{D} - \gamma_{D} - \zeta_{D} - \lambda_{D} - \phi_{D} - \psi_{D}
        \\
        \begin{split}
        \beta_{C} &= 
        \big[ \big[ \left( \Bar{\rho} \Bar{c}_{p} \Tilde{V}^{X} \right)_{e} \cdot \mathrm{Pe} \cdot \mathcal{S}_{e}, \; 0 \big] \big] 
        + \big[ \big[ \left( \Bar{\rho} \Bar{c}_{p} \Tilde{V}^{Y} \right)_{n} \cdot \mathrm{Pe} \cdot \mathcal{S}_{n}, \; 0 \big] \big] 
        + \big[ \big[ \left( \Bar{\rho} \Bar{c}_{p} \Tilde{V}^{Z} \right)_{t} \cdot \mathrm{Pe} \cdot \mathcal{S}_{t}, \; 0 \big] \big] \\
        &+ \big[ \big[ - \left( \Bar{\rho} \Bar{c}_{p} \Tilde{V}^{X} \right)_{w} \cdot \mathrm{Pe} \cdot \mathcal{S}_{w}, \; 0 \big] \big] 
        + \big[ \big[ - \left( \Bar{\rho} \Bar{c}_{p} \Tilde{V}^{Y} \right)_{s} \cdot \mathrm{Pe} \cdot \mathcal{S}_{s}, \; 0 \big] \big]
        + \big[ \big[ - \left( \Bar{\rho} \Bar{c}_{p} \Tilde{V}^{Z} \right)_{b} \cdot \mathrm{Pe} \cdot \mathcal{S}_{b}, \; 0 \big] \big]
        \end{split}
        \\
        \beta &= \beta_{D} + \beta_{C} &&
    \end{flalign}
\end{subequations}
\begin{subequations}
    \begin{flalign}
        \varphi &=  \biggl[\left( \Bar{\Tilde{Q}}_{p} \right)_{P} + \left( \Bar{\Tilde{Q}}_{cp} \right)_{P} + \left( \Bar{\Tilde{Q}}_{\Phi} \right)_{P} + \left( \Bar{\Tilde{q}}_{V} \right)_{P} \biggr] \cdot \left( \Delta \mathcal{V} \right)_{P} &&
    \end{flalign}
\end{subequations}

\newpage
\section{Transformation and Dimensionless Parameters}
\label{sec:Appendix_D}
Parameters used in the transformation of the governing equations from the Cartesian coordinate system into the normalised transformed curvilinear grid system, as well as other dimensional and dimensionless parameters relevant to the TEHL framework.

\begin{center}
  \begin{longtable}{ p{2cm} p{9cm} p{6cm} } 
    \caption{Pertinent symbols and their respective definitions and mathematical expressions.}
    \\
    \hline
    \\
    \textbf{Symbol} & \textbf{Parameter} & \textbf{Value}
    \\[1ex]
    \hline
    \\[1ex]
    $A^{i}_{j}$, $\Bar{A^{i}_{j}}$  &  Coordinate transformation and inverse coordinate transformation matrix &  Eq. \ref{eq:Coord_Transformation_Matrix}, \ref{eq:Inverse_Coord_Transformation_Matrix}
    \\[1ex]
    $A_{\scriptscriptstyle X}$  &  Normalised curvature in the $x$-direction  &  $A_{\scriptscriptstyle X} = \left( {L_x} \right)^{2} \left( {2 R_x h_0} \right)^{-1}$
    \\[1ex]
    $A_{\scriptscriptstyle Y}$  &  Normalised curvature in the $y$-direction  &  $A_{Y} = \left( {L_y} \right)^{2} \left( {2 R_y h_0} \right)^{-1}$
    \\[1ex]
    $a$  &  Hertz contact half-width of circular point contact [m] & $a = \sqrt[3] {\left( 3 W_{L} R^{'} \right) \big/ \left( 2 E^{'} \right)}$ 
    \\[1ex]
    $\mathrm{Br}$, $\mathrm{Br^{*}}$  &  Brinckman and modified Brinckman number  &  $\mathrm{Br} = \dfrac{\eta_{0} u^{2}_{e}}{k_{0} T_{0}}, \mathrm{Br^{*}} = \left( \beta_{0} T_{0} Br \right)$
    \\[1ex]
    $c_p$, $\bar{c}_{p}$  &  Dimensional $\mathrm{[J/kg.K]}$ and normalised specific heat capacity  &  $\bar{c}_{p} = {c_{p}}*\left( {c_{p_{0}}} \right)^{-1}$
    \\[1ex]
    $D^{i,j}_{k,l}$ &  Elements of influence coefficient matrix  &  - 
    \\[1ex]
    $E_{1}$, $E_{2}$ &  Lower (1) and upper (2) solid elastic modulus [Pa]  &  - 
    \\[1ex]
    $E^{'}$ &  Effective elastic modulus [Pa]  &  $E' = 2 \left[ (1-{\nu_1}^2)/E_{1}) + (1-{\nu_2}^2)/E_{2} \right] ^{-1}$ 
    \\[1ex]
    $\Bar{G}$  &  Dimensionless material parameter &  $\Bar{G} = \alpha * E^{'}$
    \\[1ex]
    $\boldsymbol{g_{i}}$, $\boldsymbol{g^{i}}$  &  Covariant and contravariant base vectors &  Eq. \ref{eq:Covariant_BaseVector}, \ref{eq:Contravariant_BaseVector} 
    \\[1ex]
    $\boldsymbol{g_{ij}}$, $\boldsymbol{g^{ij}}$  &  Covariant and contravariant metric coefficients &  Eq. \ref{eq:Metric_Coeffs} 
    \\[1ex]
    $h$, $H$  &  Dimensional [m] and normalised film thickness  &  $H = h*\left( h_{0} \right)^{-1} = \left( z_2 - z_1 \right) * \left( h_{0} \right)^{-1}$
    \\[1ex]
    $h_{0}$, $H_{0}$  &  Dimensional [m] and normalised rigid separation between two bodies prior to deformation  &  $H_{0} = {h_0} / {R} = 2.69 \Bar{U}^{0.67}  \Bar{G}^{0.53}  \Bar{W}^{-0.067}  \left( 1 - 0.61 e^{-0.73} \right)$
    \\[1ex]
    ${J}$  &  Jacobian &  Eq. \ref{eq:Coord_Transformation_Matrix}
    \\[1ex]
    $k$, $\Bar{k}$  &  Dimensional [W/m.K] and normalised thermal conductivity  &  $\bar{k} = {k}*\left( {k_0} \right)^{-1}$
    \\
    $L_{x}$, $L_{y}$  &  Characteristic length of the contact in the sliding and transverse direction, respectively  &  ($L_{x}$ = $L_{y} = a$ for circular contact)
    \\[1ex]
    $N_{x}$, $N_{y}$ &  Mesh density in the $x$- and $y$-directions, respectively  &  - 
    \\[1ex]
    $\vec{\boldsymbol{n}}$  &  Normal unit vector  &  -
    \\[1ex]
    $nb$  &  Neighbouring nodes  &  -
    \\[1ex]
    $\mathrm{Pe}$  &  Peclet number  &  $\mathrm{Pe} = \left( \rho_{0} c_{p_{0}} u_{e} L_{x} \varepsilon^{2} \right) \big/ k_{0}$
    \\[1ex]
    $\mathrm{Pr}$  &  Prandtl number  &  $\mathrm{Pr} =  \left( {\eta_{0} c_{p_{0}}} \right) \big/ k_{0}$
    \\[1ex]
    $p$, $\Bar{p}$  &  Dimensional [Pa] and normalised pressure  &  $\Bar{p} = p*\biggl[ \left( \varepsilon^{2} L_{x} \right) \big/ \left( \eta_0 u_e \right) \biggr]$
    \\[1ex]
    $p_{cav}$ &  Cavitation pressure [Pa]  &  -
    \\[1ex]
    $\Dot{q}_{V}$, $\Bar{\Dot{q}}_{V}$  &  Dimensional $\mathrm{[W/m^{3}]}$ and normalised rate of heat source/sink  &  $\Bar{\Dot{q}}_{V} = \Dot{q}_{V} \left( \varepsilon^{2} L_{x}^{2} \right) \left( k_{0} T_{0} \right)^{-1}$
    \\[1ex]
    $R'$  &  Effective radius of curvature [m]  &  ($R' = R_{x} = R_{y}$ for circular contact)
    \\[1ex]
    $R_{x}$  &  Equivalent radii of curvature of the solids (1,2) at the centre of contact in the $x$-direction [m] &  $R_{x} = {(R_{x1} R_{x2})}*{(R_{x1} + R_{x2})^{-1}}$
    \\[1ex]
    $R_{y}$  &  Equivalent radii of curvature of the solids (1,2) at the centre of contact in the $y$-direction [m] &  $R_{y} = {(R_{y1} R_{y2})}*{(R_{y1} + R_{y2})^{-1}}$
    \\[1ex]
    $r_{xy}$  &  Length aspect ratio  &  $r_{xy} = L_{x}*\left( L_{y} \right)^{-1}$
    \\[1ex]
    $SRR$  &  Slide-to-roll ratio  &  $SRR =  u_{s} \big/ u_{e}$
    \\[1ex]
    $SRR_{x}$, $SRR_{y}$  &  Ratio of sliding speed in $x$- and $y$-direction to entrainmemnt speed  &  $SRR_{x} = u_s \big/ u_e$, $SRR_{y} = v_s \big/ u_e$
    \\[1ex]
    $T$, $\Bar{T}$  &  Dimensional [K] and normalised temperature  &  $\Bar{T} = T*\left( T_{0} \right)^{-1}$
    \\[1ex]
    $T_{0}$  &  Ambient temperature [K]  &  - 
    \\[1ex]
    $t$, $\tau$  &  Dimensional [s] and normalised time  &  $\tau = t * \left( u_{e}/L_{x} \right)$
    \\[1ex]
    $U_{1}$, $V_{1}$  &  Normalised lower solid velocity in the $x$- and $y$-directions  &  $U_{1} = u_{1}/u_{e}$, $V_{1} = v_{1}/u_{e}$
    \\[1ex]
    $\Bar{U}$  &  Dimensionless speed parameter &  $\Bar{U} = u_{m} \eta_{0} * \left( E^{'} R^{'} \right)^{-1}$
    \\[1ex]
    $u$, $v$, $w$  &  Flow velocity components in the physical Cartesian space, $x$-, $y$- and $z$-directions, respectively [m/s]  &  Eq. \ref{eq:Cartesian_u}, \ref{eq:Cartesian_v}, \ref{eq:Cartesian_w}
    \\[1ex]
    $u_{e}$, $v_{e}$  &  Entertainment velocity in sliding and transverse directions [m/s]  &  $u_e = \left(u_{1} + u_{2} \right)/2$, $v_e = \left(v_{1} + v_{2} \right)/2$
    \\[1ex]
    $u_{m}(x,y)$  &  General mean velocity of surfaces [m/s], $u_{m}(x,y) = u_e$ if $u_1$ and $u_2$ are constant in space  &  $ u_m(x, y) = \bigl[u_1(x, y) + u_2(x, y)\bigr]\bigg/2$ 
    \\[1ex]
    $u_{s}$, $v_{s}$  &  Sliding velocity in sliding and transverse directions  &  $u_s = \left(u_{2} - u_{1} \right)$, $v_s = \left(v_{2} - v_{1} \right)$
    \\[1ex]
    $u_{1}$, $u_{2}$  &  Lower ($1$) and upper ($2$) solid velocity in the $x$-direction [m/s] &  -
    \\[1ex]
    $\Tilde{V}^X$, $\Tilde{V}^Y$, $\Tilde{V}^Z$  &  Contravariant velocity components in the normalised transformed domain, $X$-, $Y$- and $Z$-directions, respectively  &  Eq. \ref{eq:Ux}, \ref{eq:Vx}, \ref{eq:Wx}
    \\[1ex]
    $v_{1}$, $v_{2}$  &  Lower ($1$) and upper ($2$) solid velocity in the $y$-direction [m/s] &  -
    \\[1ex]
    $\Bar{W}$  &  Dimensionless load parameter &  $\Bar{W} = W_{L} * \left( E^{'} {R^{'}}^{2} \right)^{-1}$
    \\[1ex]
    $W_{L}$  &  Applied load $\mathrm{[N]}$  &  -
    \\[1ex]
    $x$, $X$  &  Dimensional [m] and normalised length in sliding direction  &  $X = x*\left( L_{x} \right)^{-1}$
    \\[1ex]
    $y$, $Y$  &  Dimensional [m] and normalised length in transverse direction  &  $Y = y*\left( L_{y} \right)^{-1}$
    \\[1ex]
    $Z_{1}$, $Z_{2}$  &  Lower (1) and upper (2) normalised solid height with respect to the reference origin  &  $Z_{1} = z_{1}*\left( h_{0} \right)^{-1}$, $Z_{2} = z_{2}*\left( h_{0} \right)^{-1}$
    \\[1ex]
    $z$, $Z$  &  Dimensional [m] and normalised length in film thickness direction  &  $Z = \left( z - z_{1} \right)*\left( h(x,y) \right)^{-1}$
    \\[1ex]
    $z_{1}$, $z_{2}$  &  Lower (1) and upper (2) solid height with respect to the reference origin  &  -
    \\[1ex]
    $\alpha$  &  Pressure-viscosity coefficient $\mathrm{[m^{2}/N]}$ &  - 
    \\[1ex]
    $\beta$, $\Bar{\beta}$  &  Dimensional $\mathrm{[m^{2}/N]}$ and normalised compressibility  &  $\Bar{\beta} = \beta*\left( \beta_{0} \right)^{-1}$
    \\[1ex]
    $\Dot{\gamma}$, $\Bar{\Dot{\gamma}}$  &  Dimensional $\mathrm{[s^{-1}]}$ and normalised shear-rate  &  $\Bar{\Dot{\gamma}} = \Dot{\gamma}*\biggl[ \left( h_{0} \right)/\left( u_e \right) \biggr]$
    \\[1ex]
    $\delta X_{i}$, $\delta Y_{i}$, $\delta Z_{i}$  &  Normalised length, width and height of face $i$, respectively   &  -
    \\[1ex]
    $\delta$, $\Bar{\delta}$  &  Dimensional [m] and normalised surface elastic deflection  &  $\Bar{\delta} = \delta * \left( h_0 \right)^{-1}$
    \\[1ex]
    $\varepsilon$  &  Normalised length scale  &  $\varepsilon = h_0 * \left( L_{x} \right)^{-1}$
    \\[1ex]
    $\zeta$, $\eta$, $\gamma$  &  Curvilinear coordinate system  &  -
    \\[1ex]
    $\eta$, $\bar{\eta}$  &  Dimensional [Pa.s] and normalised dynamic viscosity  &  $\bar{\eta} = {\eta}*\left( {\eta_0} \right)^{-1}$
    \\[1ex]
    $\eta_{0}$, $\rho_{0}$, $k_{0}$, $c_{p_{0}}$, $\beta_{0}$  &  Ambient dynamic viscosity [Pa.s], density $\mathrm{[kg/m^3]}$, conductivity [W/m.K], specific heat capacity $\mathrm{[J/kg.K]}$ and compressibility $\mathrm{[m^{2}/N]}$, respectively  &  -
    \\[1ex]
    $\theta$  &  Fluid film fraction  &  -
    \\[1ex]
    $\nu_{1}$, $\nu_{2}$ &  Lower (1) and upper (2) solid Poisson ratio  &  - 
    \\[1ex]
    $\rho$, $\bar{\rho}$  &  Dimensional $\mathrm{[kg/m^3]}$ and normalised density  &  $\bar{\rho} = {\rho}*\left( {\rho_0} \right)^{-1}$
    \\[1ex]
    $\sigma_{0}$  &  Diffusivity $\mathrm{[W.m^{2}.J^{-1}]}$  &  $\sigma_{0} =  {k_{0}} * \left( {\rho_{0} c_{p_{0}}} \right)^{-1}$
    \\[1ex]
    $\tau_{e}$  &  Equivalent shear stress $\mathrm{[Pa]}$ &  $\tau_{e} = \sqrt{\tau_{xz}^{2} + \tau_{yz}^{2}}$ 
    \\[1ex]
    $\tau_{0}$  &  Eyring shear stress $\mathrm{[Pa]}$ &  - 
    \\[1ex]
    $\omega_{min}$, $\omega_{max}$ &  Minimum and maximum PGMA relaxation factor threshold  &  - 
    \\[1ex]
    $\omega_{\psi}^{n}$, $\omega_{\psi}^{n-1}$ &  Relaxation factor at current ($n$) and preceding ($n-1$) iterative step  &  - 
    \\[1ex]
    $\mathcal{S}_f$  &  Surface vector normal to the control volume (CV) face $f$  &  -
    \\[1ex]
    $\Delta \mathcal{V}$  &  Volume of the control volume (CV)  &  $\Delta \mathcal{V} = \left( \delta X \right) \left( \delta Y \right) \left( \delta Z \right)$
    \\[1ex]
    $\boldsymbol{\vec{{\mathfrak{R}}}_{\psi}^{n}}$, $\boldsymbol{\vec{{\mathfrak{R}}}_{\psi}^{n-1}}$  &  Residue field at current ($n$) and preceding ($n-1$) iterative step  &  - 
  \end{longtable}
\end{center}



\newpage

\begin{thebibliography}{10}
\expandafter\ifx\csname url\endcsname\relax
  \def\url#1{\texttt{#1}}\fi
\expandafter\ifx\csname urlprefix\endcsname\relax\def\urlprefix{URL }\fi
\expandafter\ifx\csname href\endcsname\relax
  \def\href#1#2{#2} \def\path#1{#1}\fi

\bibitem{Blazek2015}
J.~Blazek, {Computational Fluid Dynamics: Principles and Applications},
  Butterworth-Heinemann, Oxford, England, UK, 2015.
\newblock \href {https://doi.org/10.1016/C2013-0-19038-1}
  {\path{doi:10.1016/C2013-0-19038-1}}.

\bibitem{Versteeg2007Feb}
H.~Versteeg, {An Introduction to Computational Fluid Dynamics: The Finite
  Volume Method}, Pearson, London, England, UK, 2007.

\bibitem{Lee2018Sep}
D.~Lee, Exploring elastohydrodynamic lubrication using finite-volume
  computational modelling techniques, Ph.D. thesis, Imperial College, London,
  England, UK (2018).
\newblock \href {https://doi.org/10.25560/81590} {\path{doi:10.25560/81590}}.

\bibitem{Ogawa1987}
S.~Ogawa, T.~Ishiguro, {A method for computing flow fields around moving
  bodies}, J. Comput. Phys. 69~(1) (1987) 49--68.
\newblock \href {https://doi.org/10.1016/0021-9991(87)90155-0}
  {\path{doi:10.1016/0021-9991(87)90155-0}}.

\bibitem{Yang1988}
H.~Q. Yang, K.~T. Yang, J.~R. Lloyd, Buoyant flow calculations with
  non-orthogonal curvilinear co-ordinates for vertical and horizontal
  parallelepiped enclosures, Int. J. Numer. Methods Eng. 25~(2) (1988)
  331--345.
\newblock \href {https://doi.org/10.1002/nme.1620250205}
  {\path{doi:10.1002/nme.1620250205}}.

\bibitem{Raithby1986}
G.~D. Raithby, P.~F. Galpin, J.~P. Van~Doormaal, Prediction of heat and fluid
  flow in complex geometries using general orthogonal coordinates, Numerical
  Heat Transfer 9~(2) (1986) 125--142.
\newblock \href {https://doi.org/10.1080/10407788608913469}
  {\path{doi:10.1080/10407788608913469}}.

\bibitem{Cannata2019}
G.~Cannata, C.~Petrelli, L.~Barsi, F.~Gallerano, Numerical integration of the
  contravariant integral form of the navier{\textendash}stokes equations in
  time-dependent curvilinear coordinate systems for three-dimensional free
  surface flows, Continuum Mech. Thermodyn. 31~(2) (2019) 491--519.
\newblock \href {https://doi.org/10.1007/s00161-018-0703-1}
  {\path{doi:10.1007/s00161-018-0703-1}}.

\bibitem{JARRAH2022122559}
I.~Jarrah, Rizwan-uddin, Nodal integral methods in general 2d curvilinear
  coordinates - applied to convection{\textendash}diffusion equation in domains
  discretized using quadrilateral elements, Int. J. Heat Mass Transfer 187
  (2022) 122559.
\newblock \href {https://doi.org/10.1016/j.ijheatmasstransfer.2022.122559}
  {\path{doi:10.1016/j.ijheatmasstransfer.2022.122559}}.

\bibitem{Kajishima}
T.~Kajishima, K.~Taira, {Computational Fluid Dynamics}, Springer International
  Publishing, Cham, Switzerland, 2017.

\bibitem{Yang1994}
H.~Q. Yang, S.~D. Habchi, A.~J. Przekwas, {General strong conservation
  formulation of Navier-Stokes equations in nonorthogonal curvilinear
  coordinates}, AIAA Journal 32~(5) (1994) 936--941.
\newblock \href {https://doi.org/10.2514/3.12077} {\path{doi:10.2514/3.12077}}.

\bibitem{Vinokur1974Feb}
M.~Vinokur, {Conservation equations of gasdynamics in curvilinear coordinate
  systems}, J. Comput. Phys. 14~(2) (1974) 105--125.
\newblock \href {https://doi.org/10.1016/0021-9991(74)90008-4}
  {\path{doi:10.1016/0021-9991(74)90008-4}}.

\bibitem{Liseikin2017Jun}
V.~D. Liseikin, {Grid Generation Methods (Scientific Computation)}, Springer,
  Berlin, Germany, 2017.

\bibitem{Farrashkhalvat2003Mar}
M.~Farrashkhalvat, {Basic Structured Grid Generation: With an Introduction to
  Unstructured Grid Generation}, Butterworth-Heinemann, Oxford, England, UK,
  2003.

\bibitem{Thompson1983}
J.~Thompson, A survey of grid generation techniques in computational fluid
  dynamics, in: 21st Aerospace Sciences Meeting, Vol. 1983-447, American
  Institute of Aeronautics and Astronautics, 1983, pp. 1--36.
\newblock \href {https://doi.org/10.2514/6.1983-447}
  {\path{doi:10.2514/6.1983-447}}.

\bibitem{Han1990}
T.~Han, R.~S. Paranjpe, A finite volume analysis of the thermohydrodynamic
  performance of finite journal bearings, J. Tribol. 112~(3) (1990) 557--565.
\newblock \href {https://doi.org/10.1115/1.2920293}
  {\path{doi:10.1115/1.2920293}}.

\bibitem{Roberts2013}
S.~A. Roberts, D.~R. Noble, E.~M. Benner, P.~R. Schunk, Multiphase hydrodynamic
  lubrication flow using a three-dimensional shell finite element model,
  Comput. Fluids 87 (2013) 12--25.
\newblock \href {https://doi.org/10.1016/j.compfluid.2012.08.009}
  {\path{doi:10.1016/j.compfluid.2012.08.009}}.

\bibitem{Moukalled}
F.~Moukalled, L.~Mangani, M.~Darwish, {The Finite Volume Method in
  Computational Fluid Dynamics}, Springer International Publishing, Cham,
  Switzerland, 2016.

\bibitem{Dowson1962Mar}
D.~Dowson, A generalized reynolds equation for fluid-film lubrication,
  International Journal of Mechanical Sciences 4~(2) (1962) 159--170.
\newblock \href {https://doi.org/10.1016/S0020-7403(62)80038-1}
  {\path{doi:10.1016/S0020-7403(62)80038-1}}.

\bibitem{Elrod1975}
H.~Elrod, M.~Adams, A computer program for cavitation and starvation problems,
  in: Leeds-Lyon Symposium on Tribology, Vol.~1, 1975, pp. 37--41.

\bibitem{Elrod1981}
H.~G. Elrod, A cavitation algorithm, J. Lubr. Technol. 103~(3) (1981) 350--354.
\newblock \href {https://doi.org/10.1115/1.3251669}
  {\path{doi:10.1115/1.3251669}}.

\bibitem{Jakobsson1965}
B.~Jakobsson, The finite journal bearing considering vaporization, Trans.
  Chalmers Univ. of Tech, Sweden 190 (1965).

\bibitem{Floberg1965}
L.~Floberg, On hydrodynamic lubricationwith special reference to sub-cavity
  pressures and number of streamers in cavitation regions, Acta Polytechnica
  Scandinavica Mechanical Engineering Series 19 (1965).

\bibitem{Floberg1973tensile}
L.~Floberg, On the Tensile Strength of Liquids, Transactions of Machine
  Elements Division / Lund Technical University, Lund, Sweden, Lund Technical
  University, 1973.

\bibitem{Ausas2009Jul}
R.~F. Ausas, M.~Jai, G.~C. Buscaglia, A mass-conserving algorithm for dynamical
  lubrication problems with cavitation, J. Tribol. 131~(3) (2009).
\newblock \href {https://doi.org/10.1115/1.3142903}
  {\path{doi:10.1115/1.3142903}}.

\bibitem{Miraskari2017May}
M.~Miraskari, F.~Hemmati, A.~Jalali, M.~Y. Alqaradawi, M.~S. Gadala, A robust
  modification to the universal cavitation algorithm in journal bearings, J.
  Tribol. 139~(3) (2017).
\newblock \href {https://doi.org/10.1115/1.4034244}
  {\path{doi:10.1115/1.4034244}}.

\bibitem{Profito2015Oct}
F.~J. Profito, M.~Giacopini, D.~C. Zachariadis, D.~Dini, A general finite
  volume method for the solution of the reynolds lubrication equation with a
  mass-conserving cavitation model, Tribology Letters 60~(18) (2015) 1--21.
\newblock \href {https://doi.org/10.1007/s11249-015-0588-0}
  {\path{doi:10.1007/s11249-015-0588-0}}.

\bibitem{Zhu2019Dec}
D.~Zhu, J.~Wang, {Interfacial mechanics: theories and methods for contact and
  Lubrication}, CRC Press, Boca Raton, FL, USA, 2019.

\bibitem{Habchi2018Jul}
W.~Habchi, {Finite Element Modeling of Elastohydrodynamic Lubrication
  Problems}, Wiley, Hoboken, NJ, USA, 2018.

\bibitem{Ferziger2020}
J.~H. Ferziger, M.~Peri{\ifmmode\acute{c}\else\'{c}\fi}, R.~L. Street,
  {Computational Methods for Fluid Dynamics}, Springer International
  Publishing, Cham, Switzerland, 2020.

\bibitem{Chung2002Feb}
T.~J. Chung, {Computational fluid dynamics}, Cambridge University Press,
  Cambridge, England, UK, 2002.
\newblock \href {https://doi.org/10.1017/CBO9780511606205}
  {\path{doi:10.1017/CBO9780511606205}}.

\bibitem{Demirdzic1982}
I.~Demirdzic, A finite volume method for computation of flow in complex
  geometries, Ph.D. thesis, Imperial College London, UK (1982).

\bibitem{Peric1982}
M.~Peric, A finite volume method for the prediction of three-dimensional fluid
  flow in complex ducts, Ph.D. thesis, Imperial College London, UK (1985).

\bibitem{Drikakis}
D.~Drikakis, W.~Rider, {High-Resolution Methods for Incompressible and
  Low-Speed Flows}, Springer, Berlin, Germany, 2005.

\bibitem{Ai1993}
X.~Ai, Numerical analyses of elastohydrodynamically lubricated line and point
  contacts with rough surfaces by using semi-system and multigrid methods
  (volumes 1 and 2), Ph.D. thesis, Northwestern University, US (1993).

\bibitem{Wang2020Aug}
Q.~J. Wang, L.~Sun, X.~Zhang, S.~Liu, D.~Zhu, {FFT-Based methods for
  computational contact mechanics}, Frontiers in Mechanical Engineering 6 (2020).
\newblock \href {https://doi.org/10.3389/fmech.2020.00061}
  {\path{doi:10.3389/fmech.2020.00061}}.

\bibitem{Hough1929Jan}
L.~A.~E. Hough, The stress produced in a semi-infinite solid by pressure on
  part of the boundary, Philosophical Transactions of the Royal Society of
  London. Series A, Containing Papers of a Mathematical or Physical Character
  228~(659-669) (1929) 377--420.
\newblock \href {https://doi.org/10.1098/rsta.1929.0009}
  {\path{doi:10.1098/rsta.1929.0009}}.

\bibitem{Ardah2023Jan}
S.~Ardah, F.~J. Profito, D.~Dini, {An integrated finite volume framework for
  thermal elasto-hydrodynamic lubrication}, Tribology International 177 (2023)
  107935.
\newblock \href {https://doi.org/10.1016/j.triboint.2022.107935}
  {\path{doi:10.1016/j.triboint.2022.107935}}.

\bibitem{Liu2000Aug}
S.~Liu, Q.~Wang, G.~Liu, {A versatile method of discrete convolution and FFT
  (DC-FFT) for contact analyses}, Wear 243~(1) (2000) 101--111.
\newblock \href {https://doi.org/10.1016/S0043-1648(00)00427-0}
  {\path{doi:10.1016/S0043-1648(00)00427-0}}.

\bibitem{Verstraete2016Oct}
T.~Verstraete, S.~Scholl, {Stability analysis of partitioned methods for
  predicting conjugate heat transfer}, International Journal of Heat and Mass
  Transfer 101 (2016) 852--869.
\newblock \href {https://doi.org/10.1016/j.ijheatmasstransfer.2016.05.041}
  {\path{doi:10.1016/j.ijheatmasstransfer.2016.05.041}}.

\bibitem{Degroote2013Sep}
J.~Degroote, {Partitioned simulation of fluid-structure interaction}, Archives
  of Computational Methods in Engineering 20~(3) (2013) 185--238.
\newblock \href {https://doi.org/10.1007/s11831-013-9085-5}
  {\path{doi:10.1007/s11831-013-9085-5}}.

\bibitem{Kaneta2014Dec}
M.~Kaneta, P.~Yang, I.~Krupka, M.~Hartl, Fundamentals of thermal
  elastohydrodynamic lubrication in si3n4 and steel circular contacts,
  Proceedings of the Institution of Mechanical Engineers, Part J: Journal of
  Engineering Tribology 229~(8) (2015) 929--939.
\newblock \href {https://doi.org/10.1177/1350650114565679}
  {\path{doi:10.1177/1350650114565679}}.

\bibitem{Kaneta2022Jun}
M.~Kaneta, K.~Matsuda, H.~Nishikawa, Effects of thermal properties of contact
  materials and slide-roll ratio in elastohydrodynamic lubrication, Journal of
  Tribology 144~(6) (2022).
\newblock \href {https://doi.org/10.1115/1.4053095}
  {\path{doi:10.1115/1.4053095}}.

\bibitem{Ohno2007Feb}
N.~Ohno, High-pressure behavior of toroidal cvt fluid for automobile, Tribol.
  Int. 40~(2) (2007) 233--238.
\newblock \href {https://doi.org/10.1016/j.triboint.2005.09.015}
  {\path{doi:10.1016/j.triboint.2005.09.015}}.

\bibitem{Lu2020Jun}
J.~Lu, T.~Reddyhoff, D.~Dini, A study of thermal effects in ehl rheology and
  friction using infrared microscopy, Tribol. Int. 146 (2020) 106179.
\newblock \href {https://doi.org/10.1016/j.triboint.2020.106179}
  {\path{doi:10.1016/j.triboint.2020.106179}}.

\bibitem{Hartinger2018Oct}
M.~Hartinger, T.~Reddyhoff, Cfd modeling compared to temperature and friction
  measurements of an ehl line contact, Tribol. Int. 126 (2018) 144--152.
\newblock \href {https://doi.org/10.1016/j.triboint.2018.05.012}
  {\path{doi:10.1016/j.triboint.2018.05.012}}.

\bibitem{Bair2019Feb}
S.~Bair, S.~Flores-Torres, The viscosity of polyalphaolefins mixtures at high
  pressure and stress, J. Tribol. 141~(2) (2019).
\newblock \href {https://doi.org/10.1115/1.4041124}
  {\path{doi:10.1115/1.4041124}}.

\bibitem{Larsson2000Apr}
R.~Larsson, O.~Andersson, Lubricant thermal conductivity and heat capacity
  under high pressure, Proceedings of the Institution of Mechanical Engineers,
  Part J: Journal of Engineering Tribology 214~(4) (2000) 337--342.
\newblock \href {https://doi.org/10.1243/1350650001543223}
  {\path{doi:10.1243/1350650001543223}}.

\bibitem{Nguyen-Schafer}
H.~Nguyen-Sch{\ifmmode\ddot{a}\else\"{a}\fi}fer, J.-P. Schmidt, {Tensor
  Analysis and Elementary Differential Geometry for Physicists and Engineers},
  Springer, Berlin, Germany, 2017.

\bibitem{Daiguji1993}
H.~Daiguji, B.~R. Shin, Some numerical schemes using curvilinear coordinate
  grids for incompressible and compressible navier-stokes equations, Sadhana
  18~(3) (1993) 431--476.
\newblock \href {https://doi.org/10.1007/BF02744365}
  {\path{doi:10.1007/BF02744365}}.

\bibitem{White2011Mar}
F.~White, Viscous Fluid Flow, 3rd Edition, McGraw-Hill, New York, US, 2007.

\bibitem{Patankar1980Jan}
S.~Patankar, Numerical Heat Transfer and Fluid Flow, CRC Press, Boca Raton, FL,
  USA, 1980.

\end{thebibliography}
\end{document}